\documentclass[english,15pt]{article}
\usepackage[T1]{fontenc}
\usepackage[latin1]{inputenc}
\usepackage{geometry}
\usepackage{float}
\usepackage{mathrsfs}
\usepackage{amsmath}
\usepackage{amssymb}
\usepackage{graphicx}
\usepackage{hyperref}
\usepackage[color=yellow]{todonotes}
\hypersetup{
    colorlinks=true,
    linkcolor=blue,
    filecolor=magenta,      
    urlcolor=cyan,
    citecolor = blue,
}

\usepackage{yhmath}
\makeatletter

\usepackage{bm}
\numberwithin{equation}{section}

\floatstyle{ruled}
\newfloat{algorithm}{tbp}{loa}
\providecommand{\algorithmname}{Algorithm}
\floatname{algorithm}{\protect\algorithmname}

\usepackage{algorithm}
 \usepackage{algorithmicx}

\usepackage{amsthm}

\usepackage{mathrsfs}
\usepackage{amssymb}

\usepackage{amsfonts}

\usepackage{epsfig}

\usepackage{bm}

\usepackage{mathrsfs}

\usepackage{enumerate}

\@ifundefined{definecolor}{\@ifundefined{definecolor}
 {\@ifundefined{definecolor}
 {\usepackage{color}}{}
}{}
}{}

\usepackage{subfig}\usepackage[all]{xy}

\newcounter{hypA}

\newcounter{hypB}

\newcounter{hypD}
\newenvironment{hypD}{\refstepcounter{hypD}\begin{itemize}
 \item[({\bf D\arabic{hypD}})]}{\end{itemize}}

\usepackage{babel}\date{}

\graphicspath{{real_data/plot/}}

\makeatother

\begin{document}

%+Title
\begin{center}

{\Large \textbf{Bayesian Inference for Non-Synchronously Observed Diffusions}}

\vspace{0.5cm}

AJAY JASRA$^{1}$,  KENGO KAMATANI$^{2}$ \& AMIN WU$^{3}$

{\footnotesize 
$^{1}$School of Data Science, The Chinese University of Hong Kong, Shenzhen, Shenzhen, CN.\\
$^{2}$Institute of Statistical Mathematics, Tokyo, 190-0014, JP. \\
$^{3}$Statistics Program,  Computer, Electrical and Mathematical Sciences and Engineering Division, King Abdullah University of Science and Technology, Thuwal, 23955-6900, KSA.}\\
{\footnotesize E-Mail:\,} \texttt{\emph{\footnotesize ajayjasra@cuhk.edu.cn, kamatani@ism.ac.jp,amin.wu@kaust.edu.sa}}

\begin{abstract}
We consider the problem of Bayesian inference for bi-variate data observed in time but with observation times
which occur non-synchronously.  In particular,  this occurs in a wide variety of applications in finance,  such as high-frequency trading or crude oil futures trading.  We adopt a diffusion model for the data and formulate a Bayesian model with priors on unknown parameters along with a latent representation for the the so-called missing data.  We then consider computational methodology to fit the model using Markov chain Monte Carlo (MCMC).
We have to resort to time-discretization methods as the complete data likelihood is intractable and this can cause considerable issues for MCMC when the data are observed in low frequencies.  In a high frequency observation frequencies we present a simple particle MCMC method based on an Euler--Maruyama time discretization,  which can be enhanced using multilevel Monte Carlo (MLMC).  In the low frequency observation  regime we introduce a novel bridging representation of the posterior in continuous time to deal with the issues of MCMC in this case.  This representation is discretized and fitted using MCMC and  MLMC.   We apply our methodology to real and simulated data to establish the efficacy of our methodology.
\end{abstract}

\end{center}

\section{Introduction}

We consider bi-variate data that are observed in time,  but the observation times of the respective time series may not coincide,  which we term the non-synchronous observation regime.  This problem has received a large amount attention in the economterics,  finance and statistics literature; see for instance the papers \cite{hay1,hay2,ogihara,mira}. The non-synchronous nature of the observations can appear in wide variety of applications,  for instance high-frequency intraday trading,  where prices of pairs of stocks are recorded in a tick-by-tick fashion (i.e.~at different times); see \cite{ait}. Other applications include crude oil futures markets see \cite{luo} for instance.

The specific type of model that we consider is the case where the data are believed to follow a diffusion process.
This presents many challenges as we now describe.  As is well-known (e.g.~\cite{kloeden}) most diffusion processes do not have an analytic form for the transition density,  assuming one exists.  Even if one has access to the transition density,  one still has data that are missing at the times where observations for the two components are not recorded; so this is at the very least a missing data problem. However,  in practice one has to time-discretize the diffusion and in a Bayesian context,  which is the case that we follow,  place a prior on the unknown parameters.

Bayesian inference for diffusion models under time discretization has received a large amount of attention from the literature; see \cite{chada,golightly,graham,jasra,papas,roberts,ryder} for a non-exhaustive list.  In most cases of practical interest,  one has to use a Markov chain Monte Carlo (MCMC) algorithm to sample from the joint posterior on the parameters and missing data of which the latter is generally the time discretization.  This problem is notoriously challenging with several articles written on how MCMC can be improved in such scenarios; see for instance in \cite{papas, roberts}.   In this paper we focus on using the particle MCMC method (PMCMC) in \cite{andrieu} and its enhancement using multilevel Monte Carlo (MLMC) \cite{chada,jasra,ml_cont,med}.  MLMC \cite{giles,giles1} enhancement of MCMC generally allows one to reduce the computational effort of MCMC to achieve a pre-specified mean square error (MSE); see \cite{ml_rev} for a review.

In the context of using PMCMC with time-discretized diffusions we consider two cases of observation regimes;
high and low frequencies.  By low frequency we mean that observations occur at $\mathcal{O}(1)$ times, whereas the high frequency observation case data are seen at arbitrarily small times. In the latter case, the well-known PMCMC and MLMC methods that have been developed seemingly work well within their constraints.  More challenging is the low frequency case.  In this case,  and has been realized in the literature previously (e.g.~\cite{beskos}) that the algorithm we use suffers as the discretization becomes increasingly more precise. In more details, for PMCMC one relies on an approximation of the likelihood function and the variance of said approximation increases exponentially fast as the time discretization goes to zero; this essentially renders the PMCMC method unreliable without an exponential effort,  associated to the discretization.  We develop a new PMCMC procedure for the case of bi-variate diffusions associated to non-synchronous observations in true continuous-time using the idea of bridges; see \cite{schauer,vd_meulen_guided_mcmc} for example.  The model is then time-discretized leading to a PMCMC kernel that has a true infinite-dimensional limit (i.e.~as the time discretization goes to zero) and subsequently solves the exponential variance issue alluded to previously.  The method is then extended to the context of multilevel PMCMC to reduce the computational effort of MCMC to achieve a pre-specified MSE.
Note that the bridging representation also facilitates a backward sampling method for PMCMC \cite{whiteley} which is known to be more efficient than regular PMCMC; although this is not implemented,  it is an important remark (and also realized in \cite{stanton} in a different context).

To summarize the contributions of this paper are as follows:
\begin{itemize}
\item{Develop fully Bayesian modeling for bi-variate diffusions associated to non-synchronous observations.}
\item{Develop MCMC and multilevel MCMC for fitting the models.}
\item{Provide conjectures for optimizing the multilevel MCMC method.}
\item{Investigate the algorithms and modeling for several simulated and real data sets.}
\end{itemize}
We note that in terms of Bayesian inference for our context, there does not seem to be a substantial amount of work; see e.g.~\cite{mira} for one contribution.

This paper is structured as follows.  
In Section \ref{sec:prob_form} we provide a more formal coverage of the problem at hand.
In Section \ref{sec:comp_app} we present our computational approaches for performing Bayesian statistical inference.
In Section \ref{sec:numerics} we give extensive simulations verifying the performance of our computational methods as well as investigating real data applications.

\section{Problem Formulation}\label{sec:prob_form}

We consider the following bi-variate diffusion process on $[0,T]$, $X_0=x_0\in\mathbb{R}^2$:
\begin{equation}\label{eq:diff}
dX_t = \mu_{\theta}(X_t)dt + \Sigma_{\theta}(X_t)dW_t
\end{equation}
where $\mu:\Theta\times\mathbb{R}^2\rightarrow\mathbb{R}^2$, $\Sigma:\Theta\times\mathbb{R}^2\rightarrow\mathbb{R}^{2\times 2}$, $\Theta\subseteq\mathbb{R}^{d}$ and $\{W_t\}_{t\in[0,T]}$ is a standard
2-dimensional Brownian motion. We assume $x_0$ is known and fixed. Set $a_{\theta}(x)=\Sigma_{\theta}(x)\Sigma_{\theta}(x)^{\top}$.
To minimize technical difficulties, the following assumption is made throughout the paper:
\begin{hypD}
        \label{hyp_diff:1}
        %We have that:
        \begin{itemize}
                \item[(i)] For each $\theta\in\Theta$,  each element of $\Sigma_{\theta}$, is twice continuously differentiable and globally Lipschitz.  $a_{\theta}(x)$ is uniformly elliptic.
                \item[(ii)] For each $\theta$,  each element of $\mu_{\theta}$ is twice continuously differentiable and globally Lipschitz.
        \end{itemize}
\end{hypD}

We have access to observations of the first dimension, $(x^1_{\tau_1^1},\dots,x^1_{\tau_{n_1}^1})$, where $0<\tau^1_1<\cdots<\tau_{n_1}^1<T$ and to observations of the second dimension, $(x^2_{\tau_1^2},\dots,x^2_{\tau_{n_2}^2})$, where $0<\tau^2_1<\cdots<\tau_{n_2}^2< T$. There is no constraint that the first collection of observation times does not include a subset of the second collection. We denote the ordered collection of all observation times as $(t_1,\dots,t_n)$ where $1\leq n\leq n_1+n_2$; typically one would have $n=n_1+n_2$.
We write the observed data $y:=(x^1_{\tau_1^1},\dots,x^1_{\tau_{n_1}^1},x^2_{\tau_1^2},\dots,x^2_{\tau_{n_2}^2})$.  For $j\in\{1,2\}$
denote by $(\nu_1^j,\dots,\nu_{m_j}^j)$ those times $(\tau_1^j,\dots,\tau_{n_j}^j)$ that the 
$j^{th}-$component is not observed; for instance one could have $(\nu_1^1,\dots,\nu_{m_1}^1)=(\tau_1^2,\dots,\tau_{n_2}^2)$ with $m_1=n_2$.
The data that are missing are denoted as $z:=(x^1_{\nu_1^1},\dots,x^1_{\nu_{m_1}^1},x_T^1,x^2_{\nu_1^2},\dots,x^2_{\nu_{m_2}^2},x_T^2)$. Assuming that it exists, the joint density of the observed and missing data is
$$
p_{\theta}(y,z) = \prod_{i=1}^{n+1}f_{\theta}(x_{t_i}|x_{t_{i-1}})
$$
where $t_0=0$,  $t_{n+1}=T$ where $f_{\theta}$ is the transition density associated to the diffusion process.

The objective is to infer the posterior distribution, with density
\begin{equation}\label{eq:post}
\pi(\theta,z|y) \propto p_{\theta}(y,z)\pi(\theta)
\end{equation}
where $\pi(\theta)$ is the prior density.

\subsection{Standard Euler--Maruyama}\label{sec:em_disc}

In practice,  we may not be able to evaluate $f_{\theta}$ nor any unbiased estimate of it.  Therefore,  one way to proceed is to consider a time discretization. Let $i\in\{1,\dots,n+1\}$
be fixed and define $\Delta_{i,l}:=(t_i-t_{i-1})2^{-l}$, $l\in\mathbb{N}_0$. Then we shall approximate $f_{\theta}(x_{t_i}|x_{t_{i-1}})$ with
$$
\int \left\{\prod_{j=1}^{2^l} f^l_{\theta}(u_{t_{i-1}+j\Delta_{i,l}}|u_{t_{i-1}+(j-1)\Delta_{i,l}})\right\} du_{t_{i-1}+\Delta_{i,l}:t_i-\Delta_{i,l}}
$$
where $u_{t_i}=x_{t_i}$, $u_{t_{i-1}}=x_{t_{i-1}}$ and $f^l$ is the Gaussian density of a $\Delta_{i,l}$ time step discretization of the diffusion process (i.e.~induced by the Euler--Maruyama discretization).  Denoting the further augmented data $z^l=(z,u_{\Delta_{1,l}:t_1-\Delta_{1,l}},\dots,u_{t_n+\Delta_{n+1,l}:t_{n+1}-\Delta_{n+1,l}})$, our objective is to sample from the density
\begin{equation}\label{eq:euler_post}
\pi^l(\theta,z^l|y) \propto \left(\prod_{i=1}^{n+1}\left\{\prod_{j=1}^{2^l} f^l_{\theta}(u_{t_{i-1}+j\Delta_{i,l}}|u_{t_{i-1}+(j-1)\Delta_{i,l}})\right\}\right)\pi(\theta).
\end{equation}
To our knowledge, the only other work on this is \cite{mira} which assumes zero drift and constant diffusion coefficient.

\subsection{Approach with Diffusion Bridges}

It is well-known that using the standard Euler--Maruyama method for Bayesian parameter estimation as in \eqref{eq:euler_post} often does not perform well when using computational methods; see e.g.~\cite{roberts}. 
The scenario where this occurs is when the observation data are observed at low frequency,  for instance at
$\mathcal{O}(1)$ times.
We introduce an approach based upon diffusion bridges which needs an initial review.

\subsubsection{Review}\label{sec:bridge_rev}

We review the method in \cite{schauer,vd_meulen_guided_mcmc} as was adopted in \cite{beskos}.
We consider the case $t\in[s_1,s_2]$, $0\leq s_1<s_2\leq T$, and let
$\mathbf{X}_{[s_1,s_2]}:=\{X_{t}\}_{t\in[s_1,s_2]}$, and $\mathbf{W}_{[s_1,s_2]}:=\{W_{t}\}_{t\in[s_1,s_2]}$.  In addition,  set $f_{\theta,t,s_{2}}(x'|x)$ denote the unknown transition density from time $t$ to $s_2$ associated to \eqref{eq:diff}; the time subscripts are added as this will prove useful below.
If one could sample from $f_{\theta,s_1,s_2}$ to obtain $(x, x')\in \mathbb{R}^{4}$.
Then
we will explain that 
we can interpolate these points by using a bridge process
which has a
a drift given by $\mu_{\theta}(x)+a_{\theta}(x)\nabla_x\log{f}_{\theta,t,s_2}(x'|x)$.
Let $\mathbb{P}_{\theta,x,x'}$ denote the law of the solution of the SDE \eqref{eq:diff}, on $[s_1,s_2]$, started at $x$ and conditioned to hit $x'$ at time $s_2$.

To continue, we introduce a user-specified auxiliary process $\{\tilde{X}_t\}_{t\in[s_1,s_2]}$ following:
\begin{align}
\label{eq:aux_SDE_tilde}
d \tilde X_{t} = \tilde \mu_{\theta}(t,\tilde X_{t})dt + \tilde \Sigma_{\theta}(t,\tilde X_{t})dW_{t}, \quad t\in[s_1,s_2],\quad~\tilde{X}_{s_1} =x, 
\end{align}
where for each $\theta\in\Theta$, $\tilde \mu_{\theta}:[s_1,s_2]\times\mathbb{R}^{2}\rightarrow\mathbb{R}^{2}$ and $\tilde \Sigma_{\theta}:\mathbb{R}^{2}\rightarrow\mathbb{R}^{2\times 2}$ is
such that for each $\theta\in\Theta$
$$
\tilde a_{\theta} (s_2,x'):= \tilde \Sigma_{\theta}(s_2,x') \tilde \Sigma_{\theta}(s_2,x')^{\top} \equiv a_{\theta}(x').
$$ 
\eqref{eq:aux_SDE_tilde} is specified so that its transition density $\tilde{f}_\theta$ is available; see \cite[Section 2.2]{schauer} for the technical conditions on $\tilde \mu_{\theta}, \tilde a_{\theta}, \tilde f_\theta$. 
The main purpose of $\{\tilde X_t\}_{t\in[s_1,s_2]}$ is to sample $x'$ and use its transition density to construct another process $\{X_t^\circ\}_{t\in[s_1,s_2]}$ conditioned to hit $x'$ at $t=s_2$; which in turn will be an importance proposal for $\{X_t\}_{t\in[s_1,s_2]}$. Let:
\begin{align}
\label{eq:aux_SDE}
d X^\circ_{t} = \mu_{\theta,s_2}^{\circ}(t,X^{\circ}_{t}; x')dt + \Sigma_{\theta}(X^{\circ}_{t})dW_{t}, \quad t\in[s_1,s_2],\quad~X^{\circ}_{s_1} =x, 
\end{align}
where:
$$
\mu_{\theta,s_2}^{\circ}(t,x;x')=\mu_{\theta}(x)+a_{\theta}(x)\nabla_x\log\tilde{f}_{\theta,t,s_2}(x'|x),
$$ 
and denote by
 $\mathbb{P}^\circ_{\theta,x,x'}$ the probability law of the solution of (\ref{eq:aux_SDE}). %and similarly  $\tilde{\mathbb{P}}_{\theta,x,x'}$ for \eqref{eq:aux_SDE_tilde}. 
The SDE in \eqref{eq:aux_SDE} yields: 
\begin{equation}
\mathbf{W}\rightarrow C_{\theta,s_1,s_2}(x,\mathbf{W}_{[s_1,s_2]},x'),%\kengo{C_{\theta,[s_1,s_2]}?} NOTATION WOULD BE INCONSISTENT AND CUMBERSOME - KEEP AS IT IS
\label{eq:map}
\end{equation} 
mapping the driving Wiener noise $\mathbf{W}$ to the solution of \eqref{eq:aux_SDE}, reparametering the problem from $\mathbf{X}$ to $(\mathbf{W},x')$.

\cite{schauer} prove that $\mathbb{P}_{\theta,x,x'}$ and $\mathbb{P}^\circ_{\theta,x,x'}$ are absolutely continuous w.r.t.~each other, with Radon--Nikod\'ym derivative: 
\begin{equation}
R_{\theta,s_1,s_2}(\mathbf{X}_{[s_1,s_2]}) := \frac{d\mathbb{P}_{\theta,x,x'} }{d \mathbb{P}^\circ_{\theta,x,x'} }(\mathbf{X}_{[s_1,s_2]})=
\exp\Big\{  \int_{s_1}^{s_2} L_{\theta,s_2}(t,X_t)dt\Big\} \times \frac{ \tilde{f}_{\theta,s_1,s_2}(x'|x)}{f_{\theta,s_1,s_2}(x'|x)},
\label{eq:density}
\end{equation}
where: 
\begin{align*}
L_{\theta,s_2}&(t,x):=\left(\mu_\theta(x)- \tilde{\mu}_\theta(t,x)\right)^{\top}\, \nabla_x\log\tilde{f}_{\theta,t,s_2}(x'|x)\\
&-\frac{1}{2}\textrm{Tr}\,\Big\{\,\big[ a_{\theta}(x)-\tilde{a}_{\theta}(t,x)\big] \big[ -\nabla_x^2\log\tilde{f}_{\theta,t,s_2}(x'|x)-\nabla_x\log\tilde{f}_{\theta,t,s_2}(x'|x)\nabla_x\log\tilde{f}_{\theta,t,s_2}(x'|x)^{\top} \big]\,\Big\} ,         
\end{align*}
with $\textrm{Tr}(\cdot)$ denoting the trace of a squared matrix.

%Althought the review here focusses upon the time interval $[0,1]$ it can easily be generalized to any
%interval $[s,t]$, $0<s<t<T$, say and to mutiple non-overlapping intervals; see e.g.~\cite{beskos}.

%Note that, in the case when $\sigma=\sigma(x)$ is not a constant function,  then, typically, $x'\rightarrow \tilde{p}_{\theta}(x,x')$ will not integrate to $1$ and will give rise to a non-trivial distribution to sample from.
%As the complete algorithm will require being able to sample from the transition density, we rewrite:
%%
%\begin{equation}
%\frac{d\overline{\mathbb{P}}_{\theta,x,x'} }{d\widetilde{\mathbb{P}}_{\theta,x,x'} }(\mathbf{X})=
%\exp\Big\{  \int_{0}^{1} L_{\theta}(t,X_t)dt\Big\} \times \frac{\tilde{p}_{\theta}(x,x')}{{p}_{\theta}(x,x')\hat{{p}}_{\theta}(x,x')}\times 
%\hat{{p}}_{\theta}(x,x'),
%\label{eq:density2}
%\end{equation}
%where an arbitrary, tractable and easy to sample density $\hat{{p}}_{\theta}(x,x')$ is used to sample $x'$.

\subsubsection{Application to Non-Synchronously Observed Diffusions}

The significance of the previous section is that it allows one to define an equivalent posterior measure, which is amenable to the computational methodology we have in mind.  Indeed, it will be possible to define algorithms in continuous time, which whilst un-implementable, will have time-discretized versions.  These latter algorithms will not suffer from the same issues as adopting the Euler--Maruyama method as in Section \ref{sec:em_disc}.

Using the construction detailed in Section \ref{sec:bridge_rev} one can consider posterior inference associated to the measure
\begin{equation}\label{eq:post_cts}
\pi^C\left(d(\mathbf{W}_{[0,T]},\theta,z)|y\right) \propto \left\{\prod_{i=1}^{n+1} R_{\theta,[t_{i-1},t_i]}\left(
C_{\theta,[t_{i-1},t_i]}(x_{t_{i-1}},\mathbf{W}_{[t_{i-1},t_i]},x_{t_i})\right)f_{\theta}(x_{t_i}|x_{t_{i-1}})
\right\}
\pi(\theta) \mathbb{P}(d\mathbf{W}_{[0,T]})d\theta dz
\end{equation}
where $d\theta$ is $d-$dimensional Lebesgue measure, $dz$ is $m:=m_1+m_2+2-$dimensional Lebesgue measure and
$\mathbb{P}$ is the law of Brownian motion on $[0,T]$. We note that $f_{\theta}(x_{t_i}|x_{t_{i-1}})=f_{\theta,[t_{i-1},t_i]}(x_{t_i}|x_{t_{i-1}})$ and so we need not evaluate this quantity (recall the defintion of $R_{\theta,[t_{i-1},t_i]}\left(
C_{\theta,[t_{i-1},t_i]}(x_{t_{i-1}},\mathbf{W}_{[t_{i-1},t_i]},x_{t_i})\right)$ in \eqref{eq:density}).

As a result of \eqref{eq:density} posterior inference on $(\theta,z)$ is equivalent under the original posterior \eqref{eq:post} and the new posterior \eqref{eq:post_cts}. That is, let $\varphi:\Theta\times\mathbb{R}^{m}\rightarrow\mathbb{R}$ be any integrable function w.r.t.~the posterior \eqref{eq:post} then we have
$$
\int_{\Theta\times\mathbb{R}^{m}}\varphi(\theta,z)\pi(\theta,z|y)d\theta dz = 
\int_{\Theta\times\mathbb{R}^{m}\times\mathcal{C}_{[0,T]}}\varphi(\theta,z)\pi^C\left(d(\mathbf{W}_{[0,T]},\theta,z)|y\right)
$$
where $\mathcal{C}_{[0,T]}$ are the collection of continuous functions on $[0,T]$.

\section{Computational Approach}\label{sec:comp_app}

\subsection{Methodology for the Euler--Maruyama Approximation}

We begin by giving a simple particle MCMC approach.  As the discretization level $l$ increases and the approximation to the diffusion becomes more accurate, this simple approach becomes computationally inefficient
in certain contexts,  but provides motivation for our next method.
We begin with the particle filter which is given in Algorithm \ref{alg:pf_disc} and the associated particle MCMC method in Algorithm \ref{alg:pmcmc_disc}.  In Algorithm \ref{alg:pmcmc_disc}, $\mathcal{U}_{[0,1]}$ denotes the uniform distribution on $[0,1]$.

The main issue with the approach in Algorithm \ref{alg:pmcmc_disc} stems from the fact that there is no well-defined algorithm as the level of discretization grows.  In other words the limiting posterior and the time-discretized one will not be absolutely continuous in the limit,  especially when the observation regime is low frequency.
 This issue as has been noted in many works (e.g.~\cite{beskos}) and can manifest itself in many inefficiencies in the associated algorithms.  In the context of PMCMC this can mean that one needs to grow the number of particles $N$ (see Algorithm \ref{alg:pf_disc}) exponentially fast as the discretization becomes more precise and can also prevent the application of variance reduction methods such as multilevel Monte Carlo. Evidence of this can be seen in Section \ref{sec:numerics}.

\begin{algorithm}
\begin{enumerate}
\item{Input: $\theta\in\Theta$, $N\in\mathbb{N}$ the number of samples and $l\in\mathbb{N}$ the level of discretization.}
\item{Initialize: Set $k_1=k_2=1$, $x_{t_0}^i=x_0$, $i\in\{1,\dots,N\}$, $p_{\theta}^N(y)=1$ and $k=1$.}
\item{Sample: For $i\in\{1,\dots,N\}$ generate $U_{t_{k-1}+\Delta_{k,l}}^i,\dots,U_{t_k-\Delta_{k,l}}^i$ using the Euler--Maruyama method with starting point $x_{t_{k-1}}^i$.  
\begin{itemize}
\item{If $\nu^{1}_{k_1}=t_k$ then for $i\in\{1,\dots,N\}$, sample $X_{t_k}^{1,i}|x_{t_k}^{2},u_{t_{k}-\Delta_{k,l}}^i$ from the Euler dynamics.  Then for $i\in\{1,\dots,N\}$ compute the weight
$$
\omega_k^i = \frac{f_{\theta}^l(x_{t_k}^{2}|x_{t_k}^{1,i},u_{t_{k}-\Delta_{k,l}}^i)}
{\sum_{j=1}^N f_{\theta}^l(x_{t_k}^{2}|x_{t_k}^{1,j},u_{t_{k}-\Delta_{k,l}}^j)}. 
$$
Compute $p_{\theta}^N(y) = p_{\theta}^N(y)\frac{1}{N}\sum_{j=1}^N f_{\theta}^l(x_{t_k}^{2}|x_{t_k}^{1,j},u_{t_{k}-\Delta_{k,l}}^j)$ and
set $k_1=k_1+1$, $x_{t_k}^{2,i}=x_{t_k}^{2}$ and go to step 4.} %dim 1 missing
\item{if $\nu^{2}_{k_2}=t_k$ then for $i\in\{1,\dots,N\}$, sample $X_{t_k}^{2,i}|x_{t_k}^{1},u_{t_{k}-\Delta_{k,l}}^i$ from the Euler dynamics.  Then for $i\in\{1,\dots,N\}$ compute the weight
$$
\omega_k^i = \frac{f_{\theta}^l(x_{t_k}^{1}|x_{t_k}^{2,i},u_{t_{k}-\Delta_{k,l}}^i)}{\sum_{j=1}^N f_{\theta}^l(x_{t_k}^{1}|x_{t_k}^{2,j},u_{t_{k}-\Delta_{k,l}}^j)}.
$$
Compute $p_{\theta}^N(y) = p_{\theta}^N(y)\frac{1}{N}\sum_{j=1}^N f_{\theta}^l(x_{t_k}^{1}|x_{t_k}^{2,j},u_{t_{k}-\Delta_{k,l}}^j)$
and set $k_2=k_2+1$, $x_{t_k}^{1,i}=x_{t_k}^{1}$ and go to step 4.} %dim 2 missing
\item{Otherwise, for $i\in\{1,\dots,N\}$ compute the weight
$$
\omega_k^i = \frac{f_{\theta}^l(x_{t_k}|u_{t_{k}-\Delta_{k,l}}^i)}{\sum_{j=1}^N f_{\theta}^l(x_{t_k}|u_{t_{k}-\Delta_{k,l}}^j)}.
$$
Compute $p_{\theta}^N(y) = p_{\theta}^N(y)\frac{1}{N}\sum_{j=1}^N f_{\theta}^l(x_{t_k}|u_{t_{k}-\Delta_{k,l}}^j)$ and
set $x_{t_k}^{i}=x_{t_k}$ and go to step 4.
} %neither missing
\end{itemize}
}
\item{Resample: Sample with replacement from the $(x_{t_1:t_k}^1,\dots,x_{t_1:t_k}^N)$ using the weights $(\omega_k^1,\dots,\omega_k^N)$ and call the resulting samples $(x_{t_1:t_k}^1,\dots,x_{t_1:t_k}^N)$ also. Set $k+1$, if $k=n+1$ go to step 5.~otherwise go to the start of step 3.}
\item{Final Sampling: For $i\in\{1,\dots,N\}$ generate $U_{t_{k-1}+\Delta_{k,l}}^i,\dots,U_{t_k-\Delta_{k,l}}^i,X_{t_k}^i$ using the Euler--Maruyama method with starting point $x_{t_{k-1}}^i$.  For $i\in\{1,\dots,N\}$ compute the weight
$
\omega_k^i = \tfrac{1}{N}.
$
and go to step 6.}
\item{Output: Pick a single trajectory from $(x_{t_1:t_{n+1}}^1,\dots,x_{t_1:t_{n+1}}^N)$ with uniform probability and denote it $z$ and return $(p_{\theta}^N(y),z)$.}
\end{enumerate}
\caption{Particle Filter for Euler-Discretization.}
\label{alg:pf_disc}
\end{algorithm}

\begin{algorithm}
\begin{enumerate}
\item{Input: $(N,S)\in\mathbb{N}^2$ the number of particles and samples and $l\in\mathbb{N}_0$ the level of discretization.}
\item{Initialize: Sample $\theta^0$ from the prior.  Run Algorithm \ref{alg:pf_disc} with $N$ samples and the given $\theta^0,l$ returning $p_{\theta^0}^N(y)$ and $z^{l,0}$. Set $k=1$ and go to step 3.}
\item{Iterate: Propose $\theta'|\theta^{k-1}$ using a proposal $q(\cdot|\theta^{k-1})$.  Run Algorithm \ref{alg:pf_disc} with $N$ samples and the given $\theta',l$ returning $p_{\theta'}^N(y)$ and $z^{l,'}$. Compute:
$$
A^l = \min\left\{1,\frac{p_{\theta'}^N(y)\pi(\theta')q(\theta^{k-1}|\theta')}{p_{\theta^{k-1}}^N(y)\pi(\theta^{k-1})q(\theta'|\theta^{k-1})}\right\}.
$$
Generate $U\sim\mathcal{U}_{[0,1]}$ and if $U<A^l$ set $(\theta^k,z^{l,k})=(\theta',z^{l,'})$ otherwise set $(\theta^k,z^{l,k})=(\theta^{k-1},z^{l,k-1})$.  Set $k=k+1$ and if $k=S+1$,
go to step 4.~otherwise go to the start of step 3.}
\item{Output: $(\theta^0,z^{l,0}),\dots,(\theta^S,z^{l,S})$.}
\end{enumerate}
\caption{Particle MCMC for Euler-Discretization.}
\label{alg:pmcmc_disc}
\end{algorithm}

\subsection{Methodology for the Diffusion Bridge Approach}

Given the issues discussed in the previous section, we now derive a PMCMC approach for parameter estimation that can deal with the afore-mentioned problems.  We begin by giving a method in continuous-time; although it cannot be implemented, on time discretization the resulting PMCMC method will perform well, even if the discretization is quite precise.

We present the particle filter in Algorithm \ref{alg:pf_cont}.  We shall introduce some notation which should clarify what is being done.  For $j\in\{1,2\}$
$$
\mathsf{M}^j = \{\nu_1^j,\dots,\nu_{m_j}^j,n+1\}
$$
$\mathsf{T}=\{t_1,\dots, t_{n+1}\}$ and $(\mathsf{M}^j)^c=\mathsf{T}\setminus\mathsf{M}^j$. Then we define for $i\in\{1,\dots,n+1\}$
\begin{align*}
\hat{h}_{\theta}(x_{t_{i}}|x_{t_{i-1}}) & = \mathbb{I}_{\mathsf{M}^1\cap(\mathsf{M}^2)^c}(t_i)\hat{f}_{\theta}(x_{t_i}^1|x_{t_i}^2,x_{t_{i-1}}) + \mathbb{I}_{(\mathsf{M}^1)^c\cap\mathsf{M}^2}(t_i)\hat{f}_{\theta}(x_{t_i}^2|x_{t_i}^1,x_{t_{i-1}})
+ \\ & \mathbb{I}_{\mathsf{M}^1\cap\mathsf{M}^2}(t_i)\hat{f}_{\theta}(x_{t_i}|x_{t_{i-1}}) + 
\mathbb{I}_{(\mathsf{M}^1)^c\cap(\mathsf{M}^2)^c}(t_i).
\end{align*}
Note that $\hat{h}_{\theta}(x_{t_{i}}|x_{t_{i-1}})$ is not a conditional density, but have written it this way for notational convenience.
The associated PMCMC algorithm is as Algorithm  
\ref{alg:pmcmc_disc}, except one replaces the calls of Algorithm \ref{alg:pf_disc} with Algorithm \ref{alg:pf_cont} and one does not need any level of time discretization.  In Algorithm \ref{alg:pf_cont} we need a proposal density
$\hat{f}_{\theta}$ on both the first and second component of position in the diffusion. 
Note that there is no requirement that the densities $\hat{f}_{\theta}(x_{t_i}^1|x_{t_i}^2,x_{t_{i-1}})$ and $\hat{f}_{\theta}(x_{t_i}^2|x_{t_i}^1,x_{t_{i-1}})$ be consistent with $\hat{f}_{\theta}(x_{t_i}|x_{t_{i-1}})$, that is, they are derived from this density, but of course it need not be the case.
 The proposals must be chosen so it can be sampled exactly and evaluated and several choices are specified later on - this component of the algorithm will be required as we continue on in the development of our methodology.
Algorithm \ref{alg:pf_cont} is clearly un-implementable as, for instance the solution mapping $C_{\theta,[t_{k-1},t_k]}$ associated to \eqref{eq:aux_SDE} is typically intractable as is the Radon--Nikod\'ym derivative
$R_{\theta,[t_{k-1},t_k]}$ and one cannot sample a continuous path of Brownian motion on a computer.  

Th PMCMC algorithm (Algorithm  
\ref{alg:pmcmc_disc}, except one replaces the calls of Algorithm \ref{alg:pf_disc} with Algorithm \ref{alg:pf_cont})
will allow approximation of
\begin{align}\label{eq:cont_post}
\pi^C\left(d(\mathbf{W}_{[0,T]},\theta,z)|y\right) & \propto \left\{\prod_{i=1}^{n+1}
R_{\theta,[t_{i-1},t_i]}\left(C_{\theta,[t_{i-1},t_i]}(x_{t_{i-1}},\mathbf{W}_{[t_{i-1},t_i]},x_{t_i})\right)
\frac{f_{\theta}(x_{t_i}|x_{t_{i-1}})}{\hat{h}_{\theta}(x_{t_{i}}|x_{t_{i-1}})}
\right\}\times\nonumber \\ & 
\left\{\prod_{i=1}^{n+1}\hat{h}_{\theta}(x_{t_{i}}|x_{t_{i-1}})\right\}
\pi(\theta)
\mathbb{P}(d\mathbf{W}_{[0,T]})
d\theta dz.
\end{align}
We explain this point in the next section as the above target is not tractable for practical simulation.

\begin{algorithm}
\begin{enumerate}
\item{Input: $\theta\in\Theta$ and $N\in\mathbb{N}$ the number of samples.}
\item{Initialize: Set $k_1=k_2=1$, $x_{t_0}^i=x_0$, $i\in\{1,\dots,N\}$, $p_{\theta}^N(y)=1$ and $k=1$.}
\item{Sample: For $i\in\{1,\dots,N\}$ generate $\mathbf{W}_{[t_{k-1},t_k]}^i$.
\begin{itemize}
\item{If $\nu^{1}_{k_1}=t_k$ then for $i\in\{1,\dots,N\}$, sample $X_{t_k}^{1,i}|x_{t_k}^{2},x_{t_{k-1}}^i$ using $\hat{f}_{\theta}(\cdot|x_{t_k}^{2},x_{t_{k-1}}^i)$ and set $x_{t_k}^{2,i}=x_{t_k}^{2}$, $k_1=k_1+1$.} %dim 1 missing
\item{If $\nu^{2}_{k_2}=t_k$ then for $i\in\{1,\dots,N\}$, sample $X_{t_k}^{2,i}|x_{t_k}^{1},x_{t_{k-1}}^i$ using $\hat{f}_{\theta}(\cdot|x_{t_k}^{1},x_{t_{k-1}}^i)$ and set $x_{t_k}^{1,i}=x_{t_k}^{1}$, $k_2=k_2+1$.} %dim 2 missing
\item{Otherwise, for $i\in\{1,\dots,N\}$,  $x_{t_k}^{i}=x_{t_k}$.
} %neither missing
\end{itemize}
Then for $i\in\{1,\dots,N\}$ compute the weight
$$
\omega_k^i = \frac{\tilde{\omega}_k^i}{\sum_{j=1}^N \tilde{\omega}_k^j} \quad
\tilde{\omega}_k^i = \left\{R_{\theta,[t_{k-1},t_k]}\left(
C_{\theta,[t_{k-1},t_k]}(x_{t_{k-1}}^i,\mathbf{w}_{[t_{k-1},t_k]}^i,x_{t_k}^i)\right)\frac{f_{\theta}(x_{t_k}^i|x_{t_{k-1}}^i)}{\hat{h}_{\theta}(x_{t_k}^i|x_{t_{k-1}}^i)}\right\}.
$$
Compute $p_{\theta}^N(y) = p_{\theta}^N(y)\frac{1}{N}\sum_{j=1}^N \tilde{\omega}_k^j$
and go to step 4.
}
\item{Resample: If $k<n+1$ sample with replacement from the $(x_{t_1:t_k}^1,\dots,x_{t_1:t_k}^N)$ using the weights $(\omega_k^1,\dots,\omega_k^N)$ and call the resulting samples $(x_{t_1:t_k}^1,\dots,x_{t_1:t_k}^N)$ also.  Otherwise do nothing.
Set $k=k+1$, if $k=n+1$ go to step 5.~otherwise go to the start of step 3.}
\item{Final Sampling: For $i\in\{1,\dots,N\}$ generate $\mathbf{W}_{[t_{n},t_{n+1}]}^i$, then $X_{t_{n+1}}^i|x_{t_n}^i$ from $\hat{f}(\cdot|x_{t_n}^i)$ and compute the weight:
$$
\omega_{n+1}^i = \frac{\tilde{\omega}_{n+1}^i}{\sum_{j=1}^N \tilde{\omega}_{n+1}^j} \quad
\tilde{\omega}_{n+1}^i = \left\{R_{\theta,[t_{n},t_{n+1}]}\left(
C_{\theta,[t_{n},t_{n+1}]}(x_{t_{n}}^i,\mathbf{w}_{[t_{n},t_{n+1}]}^i,x_{t_{n+1}}^i)\right)\frac{f_{\theta}(x_{t_{n+1}}^i|x_{t_{n}}^i)}{\hat{f}_{\theta}(x_{t_{n+1}}^{i}|x_{t_{n}}^i)}\right\}.
$$
Compute $p_{\theta}^N(y) = p_{\theta}^N(y)\frac{1}{N}\sum_{j=1}^N 
\tilde{\omega}_{n+1}^j$  and go to step 6.
}
\item{Output: Pick a single trajectory from $(x_{t_1:t_{n+1}}^1,\dots,x_{t_1:t_{n+1}}^N)$ 
using $(\omega_{n+1}^1,\dots,\omega_{n+1}^N)$
 and denote it $z$ and return $(p_{\theta}^N(y),z)$.}
\end{enumerate}
\caption{Particle Filter Using Diffusion Bridges.}
\label{alg:pf_cont}
\end{algorithm}

\subsubsection{Time Discretization}\label{sec:time_disc_db}

We now consider how one can implement a time discretization of Algorithm \ref{alg:pf_cont} and ultimately an associated PMCMC algorithm. To that end we will introduce several objects that will be needed as we proceed forwards.

We first give the standard Euler--Maruyama time discretization of the solution to \eqref{eq:aux_SDE} (associated to a time interval $[t_{i-1},t_i]$) on a regular grid of spacing $\Delta_{i,l}$, with starting point $x_{t_{i-1}}^{\circ}$ and ending point $x_{t_i}^{\circ}$. That is for $j\in\{0,1\dots,\Delta_{i,l}^{-1}-2\}$:
\begin{align}\label{eq:disc_circ}
X_{t_{i-1}+(j+1)\Delta_{i,l}}^{\circ} & = X_{t_{i-1}+j\Delta_{i,l}}^{\circ} + \mu_{\theta,t_i}^{\circ}(t_{i-1}+j\Delta_{i,l},X^{\circ}_{t_{i-1}+j\Delta_{i,l}}; x_{t_i})\Delta_{i,l} + \nonumber \\ &
\Sigma_{\theta}(X^{\circ}_{t_{i-1}+j\Delta_{i,l}})\left[W_{t_{i-1}+(j+1)\Delta_{i,l}}-W_{t_{i-1}+j\Delta_{i,l}}\right].
\end{align}
Given $(x_{t_{i-1}}^{\circ},x_{t_{i}}^{\circ})$ and $\mathbf{W}_{[t_{i-1},t_{i}]}^l=(W_{t_{i-1}+\Delta_{i,l}}-W_{t_{i-1}},\dots,W_{t_{i}}-W_{t_{i}-\Delta_{i,l}})$ \eqref{eq:disc_circ} induces a discretized path 
$X_{t_{i-1}+\Delta_{i,l}}^{\circ},\dots,X_{t_{i}-\Delta_{i,l}}^{\circ}$ and we write such a path, including the starting and ending points with the notation
$$
C_{\theta,[t_{i-1},t_i]}^l(x_{t_{i-1}}^{\circ},\mathbf{W}_{[t_{i-1},t_{i}]}^l,x_{t_{i-1}}^{\circ}).
$$
We also need a discretization of the Radon--Nikod\'ym derivative.  Consider 
$$
\mathbf{X}_{[t_{i-1},t_{i}]}^l=(X_{t_{i-1}},X_{t_{i-1}+\Delta_{i,l}},\dots,X_{t_i})
$$ 
then we set
$$
R_{\theta,[t_{i-1},t_i]}^l(\mathbf{X}_{[t_{i-1},t_{i}]}^l) := 
\exp\Big\{ \sum_{j=0}^{\Delta_{i,l}^{-1}-1} 
L_{\theta,t_i}(t,X_{t_{i-1}+j\Delta_{i,l}})\Delta_{i,l}\Big\} \times \frac{ \tilde{f}_{\theta,t_{i-1},t_i}(X_{t_i}|X_{t_{i-1}})}{f_{\theta,t_{i-1},t_i}(X_{t_i}|X_{t_{i-1}})}.
$$
We now give the time-discretized version of Algorithm \ref{alg:pf_cont} in Algorithm \ref{alg:pf_cont_disc}.
The associated PMCMC algorithm is as Algorithm  
\ref{alg:pmcmc_disc}, except one replaces the calls of Algorithm \ref{alg:pf_disc} with Algorithm \ref{alg:pf_cont_disc}.
The main difference of Algorithm \ref{alg:pf_cont_disc} to Algorithm \ref{alg:pf_cont} is that,  as stated above, we need only simulate from the finite dimensional distribution of Brownian motion,  compute time discretized versions of the solution to \eqref{eq:aux_SDE} (i.e.~as in \eqref{eq:disc_circ}) and of the Radon--Nikod\'ym derivative.  Moreoever, as $l$ grows, we would recover Algorithm \ref{alg:pf_cont}, although clearly implementing Algorithm \ref{alg:pf_cont_disc} is only feasible for some large but fixed $l$.

The PMCMC method that we have considered (Algorithm  
\ref{alg:pmcmc_disc} replacing the calls of Algorithm \ref{alg:pf_disc} with Algorithm \ref{alg:pf_cont_disc}) targets the discretized posterior
\begin{equation}\label{eq:disc_post}
\pi^l\left(d(\mathbf{W}_{[0,T]}^l,\theta,z)|y\right) \propto \left\{\prod_{i=1}^{n+1}
R_{\theta,[t_{i-1},t_i]}^l\left(C_{\theta,[t_{i-1},t_i]}^l(x_{t_{i-1}},\mathbf{W}_{[t_{i-1},t_i]}^l,x_{t_i})\right)
f_{\theta}(x_{t_i}|x_{t_{i-1}})
\right\}\pi(\theta)
\mathbb{P}(d\mathbf{W}_{[0,T]}^l)
d\theta dz.
\end{equation}
The state-space of this posterior is $\mathsf{X}^l:=\Theta\times\mathbb{R}^m\times\mathbb{R}^{2d(l)}$ where $d(l)=\sum_{j=1}^{n+1}\Delta_{j,l}^{-1}$, where $d(l)$ represents the number of increments of Brownian motion that are used at a given level of discretization. The state-space of the posterior is then feasible to sample with a finite time algorithm, at least computationally.

In order to understand the link between the posterior in \eqref{eq:disc_post} and the algorithms we have given, 
again can write
\begin{align}\label{eq:disc_post1}
\pi^l\left(d(\mathbf{W}_{[0,T]}^l,\theta,z)|y\right) & \propto \left\{\prod_{i=1}^{n+1}
R_{\theta,[t_{i-1},t_i]}^l\left(C_{\theta,[t_{i-1},t_i]}^l(x_{t_{i-1}},\mathbf{W}_{[t_{i-1},t_i]}^l,x_{t_i})\right)
\frac{f_{\theta}(x_{t_i}|x_{t_{i-1}})}{\hat{h}_{\theta}(x_{t_{i}}|x_{t_{i-1}})}
\right\}\times\nonumber \\ & 
\left\{\prod_{i=1}^{n+1}\hat{h}_{\theta}(x_{t_{i}}|x_{t_{i-1}})\right\}
\pi(\theta)
\mathbb{P}(d\mathbf{W}_{[0,T]}^l)
d\theta dz.
\end{align}
The structure of \eqref{eq:disc_post1},  conditioning on $\theta$, is similar to a state-space model, where the left-most bracket on the R.H.S.~is the likelihood, i.e.~the un-normalized weight $\tilde{\omega}_k^i$, that is used in Algorithm \ref{alg:pf_cont_disc}. The PMCMC method then provides us with samples to approximate expectations associated to \eqref{eq:disc_post1}. One can also note the similarity with the continuous-time target \eqref{eq:cont_post}.

\begin{algorithm}
\begin{enumerate}
\item{Input: $\theta\in\Theta$, $N\in\mathbb{N}$ the number of samples and $l\in\mathbb{N}$ the level of discretization.}
\item{Initialize: Set $k_1=k_2=1$, $x_{t_0}^i=x_0$, $i\in\{1,\dots,N\}$, $p_{\theta}^N(y)=1$ and $k=1$.}
\item{Sample: For $i\in\{1,\dots,N\}$ generate $W_{t_{k-1}+\Delta_{k,l}}^i-W_{t_{k-1}}^i,\dots,W_{t_{k}}^i-W_{t_{k}-\Delta_{k,l}}^i$. 
%\kengo{$=\mathbf{w}^{i,l}_{[t_{k-1},t_k]}$? If so, please check other algorithms such as Algorithm 5. } ALREADY DEFINED IN MAIN TEXT
\begin{itemize}
\item{If $\nu^{1}_{k_1}=t_k$ then for $i\in\{1,\dots,N\}$, sample $X_{t_k}^{1,i}|x_{t_k}^{2},x_{t_{k-1}}^i$ using $\hat{f}_{\theta}(\cdot|x_{t_k}^{2},x_{t_{k-1}}^i)$ and set $x_{t_k}^{2,i}=x_{t_k}^{2}$,
set $k_1=k_1+1$.
}
\item{If $\nu^{2}_{k_2}=t_k$ then for $i\in\{1,\dots,N\}$, sample $X_{t_k}^{2,i}|x_{t_k}^{1},x_{t_{k-1}}^i$ using $\hat{f}_{\theta}(\cdot|x_{t_k}^{1},x_{t_{k-1}}^i)$ and set $x_{t_k}^{1,i}=x_{t_k}^{1}$, $k_2=k_2+1$.}
\item{Otherwise, for $i\in\{1,\dots,N\}$,  $x_{t_k}^{i}=x_{t_k}$. }
\end{itemize}
Then for $i\in\{1,\dots,N\}$ compute the weight
$$
\omega_k^i = \frac{\tilde{\omega}_k^i}{\sum_{j=1}^N \tilde{\omega}_k^j} \quad
\tilde{\omega}_k^i = \left\{R_{\theta,[t_{k-1},t_k]}^l\left(
C_{\theta,[t_{k-1},t_k]}^l(x_{t_{k-1}}^i,\mathbf{w}^{i,l}_{[t_{k-1},t_k]},x_{t_k}^i)
\right)\frac{f_{\theta}(x_{t_k}^i|x_{t_{k-1}}^i)}{\hat{h}_{\theta}(x_{t_k}^{i}|x_{t_{k-1}}^i)}\right\}.
$$
Compute $p_{\theta}^N(y) = p_{\theta}^N(y)\frac{1}{N}\sum_{j=1}^N \tilde{\omega}_k^j$, 
and go to step 4.
}
\item{Resample: If $k<n+1$ sample with replacement from the $(x_{t_1:t_k}^1,\dots,x_{t_1:t_k}^N)$ using the weights $(\omega_k^1,\dots,\omega_k^N)$ and call the resulting samples $(x_{t_1:t_k}^1,\dots,x_{t_1:t_k}^N)$ also.  Otherwise do nothing. 
Set $k+1$, if $k=n+1$ go to step 5.~otherwise go to the start of step 3.}
\item{Final Sampling: For $i\in\{1,\dots,N\}$ generate $W_{t_{n}+\Delta_{n+1,l}}^i-W_{t_{n}}^i,\dots,W_{t_{n+1}}^i-W_{t_{n+1}-\Delta_{n+1,l}}^i$, then $X_{t_{n+1}}^i|x_{t_n}^i$ from $\hat{f}(\cdot|x_{t_n}^i)$ and compute the weight:
$$
\omega_{n+1}^i = \frac{\tilde{\omega}_{n+1}^i}{\sum_{j=1}^N \tilde{\omega}_{n+1}^j} \quad
\tilde{\omega}_{n+1}^i = \left\{R_{\theta,[t_{n},t_{n+1}]}^l\left(
C_{\theta,[t_{n},t_{n+1}]}^l(x_{t_{n}}^i,\mathbf{w}^{i,l}_{[t_{n},t_{n+1}]},x_{t_{n+1}}^i)
\right)\frac{f_{\theta}(x_{t_{n+1}}^i|x_{t_{n}}^i)}{\hat{f}_{\theta}(x_{t_{n+1}}^{i}|x_{t_{n}}^i)}\right\}.
$$
Compute $p_{\theta}^N(y) = p_{\theta}^N(y)\frac{1}{N}\sum_{j=1}^N \tilde{\omega}_{n+1}^j$,
and go to step 6.
}
\item{Output: Pick a single trajectory from $(x_{t_1:t_{n+1}}^1,\dots,x_{t_1:t_{n+1}}^N)$ 
using $(\omega_{n+1}^1,\dots,\omega_{n+1}^N)$
 and denote it $z$ and return $(p_{\theta}^N(y),z)$.}
\end{enumerate}
\caption{Particle Filter Using Discretized Diffusion Bridges.}
\label{alg:pf_cont_disc}
\end{algorithm}

\subsubsection{Multilevel Approach}

As our methodology relies upon time-discretization,  a natural extension is to consider the use of the Multilevel Monte Carlo method.

We now develop an extension of the method in \cite{jasra} (see also \cite{chada}) for the context of interest in this article.  Let $L\in\mathbb{N}$ be given, then the well-known multilevel identity can be written as
\begin{align}
\int_{\mathsf{X}^L}\varphi(\theta,z)
\pi^L\left(d(\mathbf{W}_{[0,T]}^L,\theta,z)|y\right) & = 
\int_{\mathsf{X}^1}\varphi(\theta,z)
\pi^1\left(d(\mathbf{W}_{[0,T]}^1,\theta,z)|y\right) + \nonumber \\ &
\sum_{l=2}^L\left\{
\int_{\mathsf{X}^l}\varphi(\theta,z)
\pi^l\left(d(\mathbf{W}_{[0,T]}^l,\theta,z)|y\right) -
\int_{\mathsf{X}^{l-1}}\varphi(\theta,z)
\pi^{l-1}\left(d(\mathbf{W}_{[0,T]}^{l-1},\theta,z|y)\right)
\right\}.\label{eq:ml_id}
\end{align}
To approximate the R.H.S.~of \eqref{eq:ml_id}, one can focus on MCMC methods which can deal with the first term and then independently the summands (independently for each $l$).  The former task can be achieved using the MCMC method given in the previous section.  The latter requires a little more work as we will now show.

We begin with the standard coupling of Brownian motion on two discrete grids.  Given,  
$\mathbf{W}^l_{[t_{i-1},t_i]}$, we will use the mapping
\begin{align*}
\mathbf{W}^{l-1}_{[t_{i-1},t_i]} &= \mathcal{S}^l(\mathbf{W}^l_{[t_{i-1},t_i]}) \\
& = \left(W_{t_{i-1}+2\Delta_{i,l}}-W_{t_{i-1}+\Delta_{i,l}}+W_{t_{i-1}+\Delta_{i,l}}-W_{t_{i-1}},\dots,
W_{t_{i}}-W_{t_{i}-\Delta_{i,l}}+W_{t_{i}-\Delta_{i,l}}-W_{t_{i}-2\Delta_{i,l}}
\right)\\
&= \left(W_{t_{i-1}+\Delta_{i,l-1}}-W_{t_{i-1}},\dots,W_{t_{i}}-W_{t_{i}-\Delta_{i,l-1}}\right).
\end{align*}
Then we set for $i\in\{1,\dots,n+1\}$, $(x_{t_{i-1}:t_i},\bar{x}_{t_{i-1}:t_{i}})\in\mathbb{R}^{8}$:
\begin{align*}
\check{h}_{\theta}\left(x_{t_{i}},\bar{x}_{t_i}|x_{t_{i-1}},\bar{x}_{t_{i-1}}\right) & = \mathbb{I}_{\mathsf{M}^1\cap(\mathsf{M}^2)^c}(t_i)\check{f}_{\theta}(x_{t_i}^1,\bar{x}_{t_i}^1|x_{t_i}^2,x_{t_{i-1}},\bar{x}_{t_{i-1}}) + \mathbb{I}_{(\mathsf{M}^1)^c\cap\mathsf{M}^2}(t_i)\check{f}_{\theta}(x_{t_i}^2,\bar{x}_{t_i}^2|x_{t_i}^1,x_{t_{i-1}},\bar{x}_{t_{i-1}})
+ \\ & \mathbb{I}_{\mathsf{M}^1\cap\mathsf{M}^2}(t_i)\check{f}_{\theta}(x_{t_i},\bar{x}_{t_i}|x_{t_{i-1}},\bar{x}_{t_{i-1}}) + 
\mathbb{I}_{(\mathsf{M}^1)^c\cap(\mathsf{M}^2)^c}(t_i)
\end{align*}
where we have used the notation that:
\begin{itemize}
\item{$\check{f}_{\theta}(x_{t_i}^1,\bar{x}_{t_i}^1|x_{t_i}^2,x_{t_{i-1}},\bar{x}_{t_{i-1}})$ is a coupling of $\hat{f}_{\theta}(x_{t_i}^1|x_{t_i}^2,x_{t_{i-1}})$
and $\hat{f}_{\theta}(\bar{x}_{t_i}^1|x_{t_i}^2,\bar{x}_{t_{i-1}})$}
\item{$\check{f}_{\theta}(x_{t_i}^2,\bar{x}_{t_i}^2|x_{t_i}^1,x_{t_{i-1}},\bar{x}_{t_{i-1}})$ is a coupling of
$\check{f}_{\theta}(x_{t_i}^2|x_{t_i}^1,x_{t_{i-1}})$ and $\check{f}_{\theta}(\bar{x}_{t_i}^2|x_{t_i}^1,\bar{x}_{t_{i-1}})$}
\item{$\check{f}_{\theta}(x_{t_i},\bar{x}_{t_i}|x_{t_{i-1}},\bar{x}_{t_{i-1}})$
is a coupling of $\check{f}_{\theta}(x_{t_i}|x_{t_{i-1}})$
and $\check{f}_{\theta}(\bar{x}_{t_i}|\bar{x}_{t_{i-1}})$.
}
\end{itemize}
The exact nature of the afore-mentioned coupling is open, in that we need not specify one, but in this article all such couplings will be the maximal coupling.
Then we set for $i\in\{1,\dots,n+1\}$
\begin{align*}
\check{R}_{\theta,[t_{i-1},t_i]}^l\left((x_{t_{i-1}},\bar{x}_{t_{i-1}}),\mathbf{W}^l_{[t_{i-1},t_i]},(x_{t_{i}},\bar{x}_{t_{i}})\right)& :=
 \frac{1}{2}\left\{R_{\theta,[t_{i-1},t_i]}^l\left(
C_{\theta,[t_{i-1},t_i]}^l(x_{t_{i-1}},\mathbf{W}^{l}_{[t_{i-1},t_i]},x_{t_i})
\right)\frac{f_{\theta}(x_{t_i}|x_{t_{i-1}})}{\hat{h}_{\theta}(x_{t_k}|x_{t_{k-1}})}\right\} + \\ &
 \frac{1}{2}\left\{R_{\theta,[t_{i-1},t_i]}^{l-1}\left(
C_{\theta,[t_{i-1},t_i]}^{l-1}(\bar{x}_{t_{i-1}},\mathcal{S}^l(\mathbf{W}^{l}_{[t_{i-1},t_i]}),\bar{x}_{t_i})
\right)\frac{f_{\theta}(\bar{x}_{t_i}|\bar{x}_{t_{i-1}})}{\hat{h}_{\theta}(\bar{x}_{t_k}|\bar{x}_{t_{k-1}})}\right\}.
\end{align*}
Now we use the notation that $\bar{z}=(\bar{x}^1_{\nu_1^1},\dots,\bar{x}_{\nu_{m_1}^1}^1,\bar{x}_T^1,
\bar{x}^2_{\nu_1^2},\dots,\bar{x}_{\nu_{m_2}^2}^2,\bar{x}_T^2)$ and that $\bar{x}^j_{\tau_i^j}=
x^j_{\tau_i^j}$, $j\in\{1,2\}$ $i\in\{1,\dots,n_j^j\}$.
Then we define the extended target:
\begin{align}
\check{\pi}^l\left(d(\mathbf{W}_{[0,T]}^l,\theta,z,\bar{z})|y\right)& \propto \left\{\prod_{i=1}^{n+1}
\check{R}_{\theta,[t_{i-1},t_i]}^l\left((x_{t_{i-1}},\bar{x}_{t_{i-1}}),\mathbf{W}^l_{[t_{i-1},t_i]},(x_{t_{i}},\bar{x}_{t_{i}})\right)
\right\}\times\nonumber \\ & 
\left\{\prod_{i=1}^{n+1}
\check{h}_{\theta}\left(x_{t_{i}},\bar{x}_{t_i}|x_{t_{i-1}},\bar{x}_{t_{i-1}}\right)\right\}
\pi(\theta)
\mathbb{P}(d\mathbf{W}_{[0,T]}^l)
d\theta dzd\bar{z}
\end{align}
where the state-space here is $\check{\mathsf{X}}^l=\mathsf{X}^l\times\mathbb{R}^m$.

Now set
\begin{eqnarray*}
V^l(\mathbf{W}_{[0,T]}^l,\theta,z,\bar{z}) & = & \prod_{i=1}^{n+1}\frac{R_{\theta,[t_{i-1},t_i]}^l\left(C_{\theta,[t_{i-1},t_i]}^l(x_{t_{i-1}},\mathbf{W}_{[t_{i-1},t_i]}^l,x_{t_i})\right)
\frac{f_{\theta}(x_{t_i}|x_{t_{i-1}})}{\hat{h}_{\theta}(x_{t_{i}}|x_{t_{i-1}})}
}{\check{R}_{\theta,[t_{i-1},t_i]}^l\left((x_{t_{i-1}},\bar{x}_{t_{i-1}}),\mathbf{W}^l_{[t_{i-1},t_i]},(x_{t_{i}},\bar{x}_{t_{i}})\right)
}\\
\bar{V}^l(\mathbf{W}_{[0,T]}^l,\theta,z,\bar{z}) & = &
\prod_{i=1}^{n+1}\frac{R_{\theta,[t_{i-1},t_i]}^{l-1}\left(C_{\theta,[t_{i-1},t_i]}^{l-1}(\bar{x}_{t_{i-1}},\mathcal{S}^l(\mathbf{W}_{[t_{i-1},t_i]}^l),\bar{x}_{t_i})\right)
\frac{f_{\theta}(\bar{x}_{t_i}|\bar{x}_{t_{i-1}})}{\hat{h}_{\theta}(\bar{x}_{t_{i}}|\bar{x}_{t_{i-1}})}
}{\check{R}_{\theta,[t_{i-1},t_i]}^l\left((x_{t_{i-1}},\bar{x}_{t_{i-1}}),\mathbf{W}^l_{[t_{i-1},t_i]},(x_{t_{i}},
\bar{x}_{t_{i}})\right)
}
\end{eqnarray*}
Returning to the R.H.S.~of the multilevel identity \eqref{eq:ml_id} one can easily verify that
$$
\int_{\mathsf{X}^l}\varphi(\theta,z)
\pi^l\left(d(\mathbf{W}_{[0,T]}^l,\theta,z)|y\right) -
\int_{\mathsf{X}^{l-1}}\varphi(\theta,z)
\pi^{l-1}\left(d(\mathbf{W}_{[0,T]}^{l-1},\theta,z|y)\right)
 = 
$$
$$
\frac{
\int_{\check{\mathsf{X}}^l}\varphi(\theta,z)
V^l(\mathbf{W}_{[0,T]}^l,\theta,z,\bar{z})\check{\pi}^l\left(d(\mathbf{W}_{[0,T]}^l,\theta,z,\bar{z})|y\right)
}
{
\int_{\check{\mathsf{X}}^l}
V^l(\mathbf{W}_{[0,T]}^l,\theta,z,\bar{z})\check{\pi}^l\left(d(\mathbf{W}_{[0,T]}^l,\theta,z,\bar{z})|y\right)
}
-
\frac{
\int_{\check{\mathsf{X}}^l}\varphi(\theta,\bar{z})
\bar{V}^l(\mathbf{W}_{[0,T]}^l,\theta,z,\bar{z})\check{\pi}^l\left(d(\mathbf{W}_{[0,T]}^l,\theta,z,\bar{z})|y\right)
}
{
\int_{\check{\mathsf{X}}^l}
\bar{V}^l(\mathbf{W}_{[0,T]}^l,\theta,z,\bar{z})\check{\pi}^l\left(d(\mathbf{W}_{[0,T]}^l,\theta,z,\bar{z})|y\right)
}.
$$
Therefore, to approximate the multilevel identity using this idea, we require a method to sample from the target 
$\check{\pi}^l$, where $l\in\{2,\dots,L\}$. This comprises the delta particle filter from \cite{jasra} in Algorithm \ref{alg:pf_cont_delta} and the associated PMCMC method in Algorithm \ref{alg:pmcmc_cont_delta}.

To describe Algorithm \ref{alg:pf_cont_delta} we need some notation as follows
\begin{align}
\tilde{\omega}_k^{i,l} & = \frac{1}{2}\left\{R_{\theta,[t_{k-1},t_k]}^l\left(
C_{\theta,[t_{k-1},t_k]}^l(x_{t_{k-1}}^{i,l},\mathbf{w}^{i,l}_{[t_{k-1},t_k]},x_{t_k}^{i,l})
\right)\frac{f_{\theta}(x_{t_k}^{i,l}|x_{t_{k-1}}^{i,l})}{\hat{f}_{\theta}(x_{t_k}^{1,i,l}|x_{t_k}^{2},x_{t_{k-1}}^{i,l})}\right\} +\nonumber\\ &
\frac{1}{2}\left\{R_{\theta,[t_{k-1},t_k]}^{l-1}\left(
C_{\theta,[t_{k-1},t_k]}^{l-1}(\bar{x}_{t_{k-1}}^{i,l-1},\mathbf{w}^{i,l-1}_{[t_{k-1},t_k]},\bar{x}_{t_k}^{i,l-1})
\right)\frac{f_{\theta}(\bar{x}_{t_k}^{i,l-1}|\bar{x}_{t_{k-1}}^{i,l-1})}{\hat{f}_{\theta}(\bar{x}_{t_k}^{1,i,l-1}|x_{t_k}^{2},\bar{x}_{t_{k-1}}^{i,l-1})}\right\}.\label{eq:om_1}
\end{align}
In another case we have
\begin{align}
\tilde{\omega}_k^{i,l} & = \frac{1}{2}\left\{R_{\theta,[t_{k-1},t_k]}^l\left(
C_{\theta,[t_{k-1},t_k]}^l(x_{t_{k-1}}^{i,l},\mathbf{w}^{i,l}_{[t_{k-1},t_k]},x_{t_k}^{i,l})
\right)\frac{f_{\theta}(x_{t_k}^{i,l}|x_{t_{k-1}}^{i,l})}{\hat{f}_{\theta}(x_{t_k}^{2,i,l}|x_{t_k}^{1},x_{t_{k-1}}^{i,l})}\right\} +\nonumber\\ &
\frac{1}{2}\left\{R_{\theta,[t_{k-1},t_k]}^{l-1}\left(
C_{\theta,[t_{k-1},t_k]}^{l-1}(\bar{x}_{t_{k-1}}^{i,l-1},\mathbf{w}^{i,l-1}_{[t_{k-1},t_k]},\bar{x}_{t_k}^{i,l-1})
\right)\frac{f_{\theta}(\bar{x}_{t_k}^{i,l-1}|\bar{x}_{t_{k-1}}^{i,l-1})}{\hat{f}_{\theta}(\bar{x}_{t_k}^{2,i,l-1}|x_{t_k}^{1},\bar{x}_{t_{k-1}}^{i,l-1})}\right\}.\label{eq:om_2}
\end{align}
The final case:
\begin{align}
\tilde{\omega}_k^{i,l} & = \frac{1}{2}\Big\{R_{\theta,[t_{k-1},t_k]}^l\left(
C_{\theta,[t_{k-1},t_k]}^l(x_{t_{k-1}}^{i,l},\mathbf{w}^{i,l}_{[t_{k-1},t_k]},x_{t_k}^{i,l})
\right) + \nonumber\\ &
R_{\theta,[t_{k-1},t_k]}^{l-1}\left(
C_{\theta,[t_{k-1},t_k]}^{l-1}(\bar{x}_{t_{k-1}}^{i,l-1},\mathbf{w}^{i,l-1}_{[t_{k-1},t_k]},\bar{x}_{t_k}^{i,l-1})
\right)\Big\}.\label{eq:om_3}
\end{align}
In Algorithm \ref{alg:pmcmc_cont_delta}, $\check{z}=(z,\bar{z})$ is used to represent the pair of missing trajectories used at levels $l$ and $l-1$ respectively.

\begin{algorithm}
\begin{enumerate}
\item{Input: $\theta\in\Theta$, $N\in\mathbb{N}$ the number of samples and $l\in\mathbb{N}$ the level of discretization.}
\item{Initialize: Set $k_1=k_2=1$, $x_{t_0}^{i,l}=\bar{x}_{t_0}^{i,l}=x_0$, $i\in\{1,\dots,N\}$, $p_{\theta}^N(y)=1$ and $k=1$.}
\item{Sample: For $i\in\{1,\dots,N\}$ generate $W_{t_{k-1}+\Delta_{k,l}}^i-W_{t_{k-1}}^i,\dots,W_{t_{k}}^i-W_{t_{k}-\Delta_{k,l}}^i$ and sum the relevant increments to obtain
$W_{t_{k-1}+\Delta_{k,l-1}}^i-W_{t_{k-1}}^i,\dots,W_{t_{k}}^i-W_{t_{k}-\Delta_{k,l-1}}^i$.
\begin{itemize}
\item{If $\nu^{1}_{k_1}=t_k$ then for $i\in\{1,\dots,N\}$, sample $(X_{t_k}^{1,i,l},\bar{X}_{t_k}^{1,i,l-1})|x_{t_k}^{2},x_{t_{k-1}}^{i,l},\bar{x}_{t_{k-1}}^{i,l-1}$ using any coupling of $\hat{f}_{\theta}(\cdot|x_{t_k}^{2},x_{t_{k-1}}^{i,l})$ 
and $\hat{f}_{\theta}(\cdot|x_{t_k}^{2},\bar{x}_{t_{k-1}}^{i,l-1})$ 
and set $x_{t_k}^{2,i,l}=\bar{x}_{t_k}^{2,i,l-1}=x_{t_k}^{2}$. 
Then for $i\in\{1,\dots,N\}$ compute the weight
$
\omega_k^{i,l} = \tilde{\omega}_k^{i,l}\Big/\sum_{j=1}^N \tilde{\omega}_k^{j,l}
$ where $\tilde{\omega}_k^{i,l}$ is as \eqref{eq:om_1}.
Compute $p_{\theta}^N(y) = p_{\theta}^N(y)\frac{1}{N}\sum_{j=1}^N \tilde{\omega}_k^j$, 
set $k_1=k_1+1$ and go to step 4.}
\item{If $\nu^{2}_{k_2}=t_k$ then for $i\in\{1,\dots,N\}$,  sample $(X_{t_k}^{2,i,l},\bar{X}_{t_k}^{2,i,l-1})|x_{t_k}^{1},x_{t_{k-1}}^{i,l},\bar{x}_{t_{k-1}}^{i,l-1}$ using any coupling of $\hat{f}_{\theta}(\cdot|x_{t_k}^{1},x_{t_{k-1}}^{i,l})$ 
and $\hat{f}_{\theta}(\cdot|x_{t_k}^{1},\bar{x}_{t_{k-1}}^{i,l-1})$ 
and set $x_{t_k}^{1,i,l}=\bar{x}_{t_k}^{1,i,l-1}=x_{t_k}^{1}$. 
Then for $i\in\{1,\dots,N\}$ compute the weight
$
\omega_k^{i,l} = \tilde{\omega}_k^{i,l}\Big/\sum_{j=1}^N \tilde{\omega}_k^{j,l}
$ where $\tilde{\omega}_k^{i,l}$ is as \eqref{eq:om_2}.
Compute $p_{\theta}^N(y) = p_{\theta}^N(y)\frac{1}{N}\sum_{j=1}^N \tilde{\omega}_k^j$, 
set $k_2=k_2+1$ and go to step 4.}
\item{Otherwise, for $i\in\{1,\dots,N\}$,  $x_{t_k}^{i,l}=\bar{x}_{t_k}^{i,l-1}=x_{t_k}$. 
Then for $i\in\{1,\dots,N\}$ compute the weight
$
\omega_k^{i,l} = \tilde{\omega}_k^{i,l}\Big/\sum_{j=1}^N \tilde{\omega}_k^{j,l}
$ where $\tilde{\omega}_k^{i,l}$ is as \eqref{eq:om_3}.
Compute $p_{\theta}^N(y) = p_{\theta}^N(y)\frac{1}{N}\sum_{j=1}^N \tilde{\omega}_k^j$, and go to step 4.}
\end{itemize}
}
\item{Resample: If $k<n+1$ sample with replacement from the $\left((x_{t_1:t_k}^{1,l},\bar{x}_{t_1:t_k}^{1,l-1},
\mathbf{w}^{1,l}_{[0,t_k]}),\dots,(x_{t_1:t_k}^{N,l},\bar{x}_{t_1:t_k}^{N,l-1},\mathbf{w}^{N,l}_{[0,t_k]})\right)$ using the weights $(\omega_k^1,\dots,\omega_k^N)$ and call the resulting samples 
$\left((x_{t_1:t_k}^{1,l},\bar{x}_{t_1:t_k}^{1,l-1},\mathbf{w}^{1,l}_{[0,t_k]}),\dots,(x_{t_1:t_k}^{N,l},\bar{x}_{t_1:t_k}^{N,l-1},\mathbf{w}^{N,l}_{[0,t_k]})\right)$ also. 
Otherwise do nothing.
Set $k+1$, if $k=n+1$ go to step 5.~otherwise go to the start of step 3.}
\item{Final Sampling: For $i\in\{1,\dots,N\}$ generate $W_{t_{n}+\Delta_{n+1,l}}^i-W_{t_{n}}^i,\dots,W_{t_{n+1}}^i-W_{t_{n+1}-\Delta_{n+1,l}}^i$ and sum the relevant increments to obtain
$W_{t_{n}+\Delta_{n+1,l-1}}^i-W_{t_{n}}^i,\dots,W_{t_{n+1}}^i-W_{t_{n+1}-\Delta_{n+1,l-1}}^i$, then 
$(X_{t_{n+1}}^{i,l},\bar{X}_{t_{n+1}}^{i,l-1})|x_{t_{n}}^{i,l},\bar{x}_{t_{n}}^{i,l-1}$ using any coupling of $\hat{f}_{\theta}(\cdot|x_{t_{n}}^{i,l})$ 
and $\hat{f}_{\theta}(\cdot|\bar{x}_{t_{n}}^{i,l-1})$ and compute the weight $
\omega_{n+1}^{i,l} = \tilde{\omega}_{n+1}^{i,l}\Big/\sum_{j=1}^N \tilde{\omega}_{n+1}^{j,l}
$ where 
\begin{align*}
\tilde{\omega}_{n+1}^{i,l} & = \frac{1}{2}\left\{R_{\theta,[t_{n},t_{n+1}]}^l\left(
C_{\theta,[t_{n},t_{n+1}]}^l(x_{t_{n}}^{i,l},\mathbf{w}^{i,l}_{[t_{n},t_{n+1}]},x_{t_{n+1}}^{i,l})
\right)\frac{f_{\theta}(x_{t_{n+1}}^{i,l}|x_{t_{n}}^{i,l})}{\hat{f}_{\theta}(x_{t_{n+1}}^{i,l}|x_{t_{n}}^{i,l})}\right\} +\\ &
\frac{1}{2}\left\{R_{\theta,[t_{n},t_{n+1}]}^{l-1}\left(
C_{\theta,[t_{n},t_{n+1}]}^{l-1}(\bar{x}_{t_{n}}^{i,l-1},\mathbf{w}^{i,l-1}_{[t_{n},t_{n+1}]},\bar{x}_{t_{n+1}}^{i,l-1})
\right)\frac{f_{\theta}(\bar{x}_{t_{n+1}}^{i,l-1}|\bar{x}_{t_{n}}^{i,l-1})}{\hat{f}_{\theta}(\bar{x}_{t_{n+1}}^{i,l-1}|\bar{x}_{t_{n}}^{i,l-1})}\right\}.
\end{align*}
Compute $p_{\theta}^N(y) = p_{\theta}^N(y)\frac{1}{N}\sum_{j=1}^N \tilde{\omega}_{n+1}^j$, and go to step 6.
}
\item{Output: Pick a single pair of trajectories and Brownian motion from $\left((x_{t_1:t_{n+1}}^{1,l},\bar{x}_{t_1:t_{n+1}}^{1,l-1},\mathbf{w}^{1,l}_{[0,t_{n+1}]}),\dots,(x_{t_1:t_{n+1}}^{N,l},\bar{x}_{t_1:t_{n+1}}^{N,l-1},\mathbf{w}^{N,l}_{[0,t_{n+1}]})\right)$
using $(\omega_{n+1}^1,\dots,\omega_{n+1}^N)$
 and denote the trajectories as $\check{z}$ and return $(p_{\theta}^N(y),\check{z},\mathbf{w}^{l}_{[0,T]})$.}
\end{enumerate}
\caption{Delta Particle Filter Using Discretized Diffusion Bridges.}
\label{alg:pf_cont_delta}
\end{algorithm}

\begin{algorithm}
\begin{enumerate}
\item{Input: $(N,S)\in\mathbb{N}^2$ the number of particles and samples and $l\in\mathbb{N}_0$ the level of discretization.}
\item{Initialize: Sample $\theta^{0,l}$ from the prior.  Run Algorithm \ref{alg:pf_cont_delta} with $N$ samples and the given $\theta^{0,l},l$ returning $p_{\theta^{0,l}}^N(y)$ and $\check{z}^{0,l},\mathbf{w}^{0,l}_{[0,T]}$. Set $k=1$ and go to step 3.}
\item{Iterate: Propose $\theta'|\theta^{k-1,l}$ using a proposal $q(\cdot|\theta^{k-1,l})$.  Run Algorithm \ref{alg:pf_cont_delta} with $N$ samples and the given $\theta',l$ returning $p_{\theta'}^N(y)$ and $\check{z}^{',l},\mathbf{w}^{',l}_{[0,T]}$. Compute:
$$
A^l = \min\left\{1,\frac{p_{\theta'}^N(y)\pi(\theta')q(\theta^{k-1,l}|\theta')}{p_{\theta^{k-1,l}}^N(y)\pi(\theta^{k-1,l})q(\theta'|\theta^{k-1,l})}\right\}.
$$
Generate $U\sim\mathcal{U}_{[0,1]}$ and if $U<A^l$ set $(\theta^{k,l},\check{z}^{k,l},\mathbf{w}^{k,l}_{[0,T]})=(\theta',\check{z}^{',l},\mathbf{w}^{',l}_{[0,T]})$ otherwise set $(\theta^{k,l},\check{z}^{k,l},\mathbf{w}^{k,l}_{[0,T]})=(\theta^{k-1,l},\check{z}^{k-1,l},\mathbf{w}^{k-1,l}_{[0,T]})$.  Set $k=k+1$ and if $k=S+1$,
go to step 4.~otherwise go to the start of step 3.}
\item{Output: $(\theta^{0,l},\check{z}^{0,l},\mathbf{w}^{0,l}_{[0,T]}),\dots,(\theta^{S,l},\check{z}^{S,l},\mathbf{w}^{S,l}_{[0,T]})$.}
\end{enumerate}
\caption{Particle MCMC for Extended Target.}
\label{alg:pmcmc_cont_delta}
\end{algorithm}

\subsection{Final Algorithm}

We present our final multilevel MCMC method based upon discretized diffusion bridges. 
We assume that we have $L$ and number of samples $S_1,\dots,S_L$ already given; we will discuss how to choose these below. The method is then as follows.
\begin{enumerate}
\item{Run the PMCMC method to approximate $\pi^1$ for $S_1$ iterations producing $(\theta^{0,1},z^{0,1}),\dots,(\theta^{S_1,1},z^{S_1,1})$.}
\item{For $l\in\{2,\dots,L\}$ independently and independently of Step 1.~run the PMCMC method in Algorithm
\ref{alg:pmcmc_cont_delta} for $S_l$ iterations producing $(\theta^{0,l},\check{z}^{0,l},\mathbf{w}^{0,l}_{[0,T]}),\dots,(\theta^{S_l,l},\check{z}^{S_l,l},\mathbf{w}^{S_l,l}_{[0,T]})$.
}
\end{enumerate}
Set $\pi^L(\varphi) = \int_{\mathsf{X}^L}\varphi(\theta,z)
\pi^L\left(d(\mathbf{W}_{[0,T]}^L,\theta,z)|y\right)$.
The etimator we will use is then:
$$
\widehat{\pi^L(\varphi)} = \frac{1}{S_1+1}\sum_{i=0}^{S_1} \varphi(\theta^{i,1},z^{i,1}) +
$$
$$
\sum_{l=2}^L\left\{\frac{\tfrac{1}{S_l+1}\sum_{i=0}^{S_l}
V^l(\mathbf{w}_{[0,T]}^{i,l},\theta^{i,l},z^{i,l},\bar{z}^{i,l})
\varphi(\theta^{i,l},z^{i,l})}{\tfrac{1}{S_l+1}\sum_{i=0}^{S_l}
V^l(\mathbf{w}_{[0,T]}^{i,l},\theta^{i,l},z^{i,l},\bar{z}^{i,l})} - \frac{
\tfrac{1}{S_l+1}\sum_{i=0}^{S_l}
\bar{V}^l(\mathbf{w}_{[0,T]}^{i,l},\theta^{i,l},z^{i,l},\bar{z}^{i,l})
\varphi(\theta^{i,l},\bar{z}^{i,l})
}{\tfrac{1}{S_l+1}\sum_{i=0}^{S_l}
\bar{V}^l(\mathbf{w}_{[0,T]}^{i,l},\theta^{i,l},z^{i,l},\bar{z}^{i,l})}\right\}.
$$

Although we do not have a theory for considering, for instance, the MSE of this estimator, as
was done in \cite{jasra},  we do have a conjecture based on the afore-mentioned article and \cite{papas}.  As we
are essentially using the Euler--Maruyama method,  we expect that for an appropriate class of (elliptic) diffusion
processes with non-constant diffusion coefficient, that one should choose, so as to achieve a MSE of $\mathcal{O}(\epsilon^2)$ $L=\mathcal{O}(|\log(\epsilon)|)$ and $S_l=\mathcal{O}(\epsilon^{-2}2^{-l}L)$ which, ignoring $T$,
should have a cost of $\mathcal{O}(\epsilon^{-2}\log(\epsilon)^2)$.  If however, one considered only a single
level $L$, then we would expect the cost to achieve a MSE of $\mathcal{O}(\epsilon^2)$ would be $\mathcal{O}(\epsilon^{-3})$.   We note that if $\Sigma_{\theta}$ is constant in $x$ then we expect an improvement in that
choosing $L$ as above and  $S_l=\mathcal{O}(\epsilon^{-2}2^{-3l/2})$ then one has an optimal cost of 
$\mathcal{O}(\epsilon^{-2})$. These conjectures will be considered in Section \ref{sec:numerics}.

\section{Numerical Simulations}\label{sec:numerics}

\subsection{Variance Comparison: Euler--Maruyama vs. Diffusion Bridge}

We assess the estimated likelihood variance between using a Particle Filter (PF) with an Euler--Maruyama (EM) and then the Diffusion Bridge (DB) approach for a bi-variate Ornstein--Uhlenbeck (OU) process with missing components in the observations.  The subsequent results will demonstrate the limitations of using PMCMC (which relies on the likelihood estimate from the PF) for a simple EM discretization,  in the context of low-frequency data.

\begin{figure}[h!]
  \centering
  \includegraphics[width=0.8\textwidth,height=6cm]{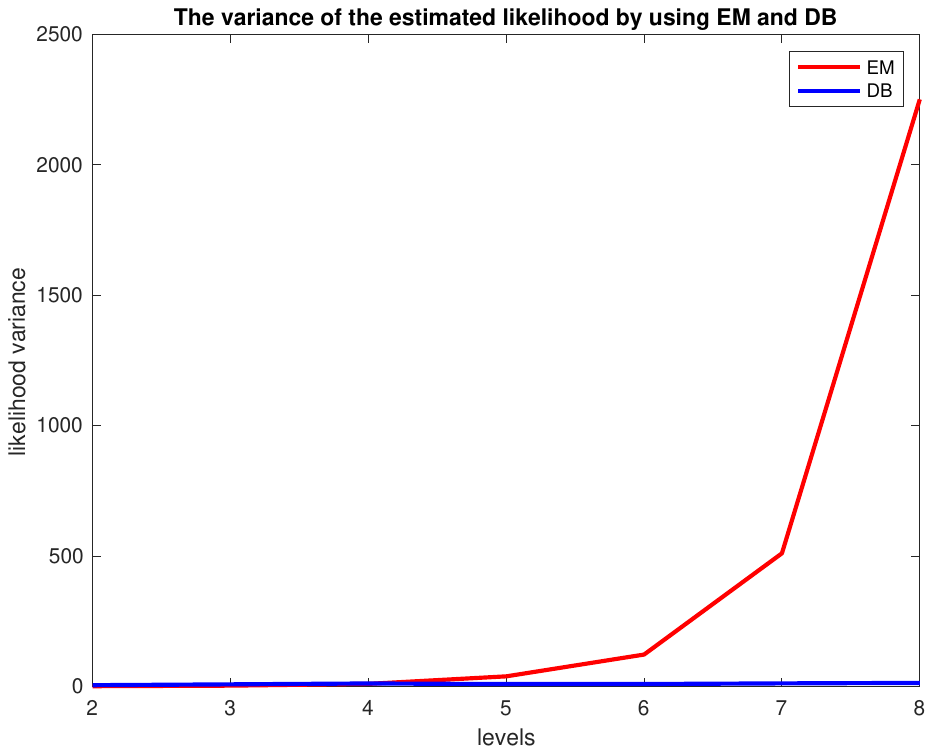}
\caption{Comparison of the Variance of the Log Likelihood Estimator using a Particle Filter. In the plot EM is red and and DB is blue.}
  \label{fig:likelihood_comp}
\end{figure}

We generate $T = 50$ data points for the following bi-variate OU process
$$
    dX_t = -AX_t dt + \Sigma dW_t
$$
where the parameter matrices are: 
$$A = 
	\begin{bmatrix}
	    0.8 & 0.2 \\ -0.3 & 0.8
	\end{bmatrix}
\qquad\textrm{and}\qquad \Sigma = 
\begin{bmatrix}
    1 & 0.5 \\ 0.5 & 1
\end{bmatrix}.
$$
In our simulation of the data,  we omit $13$ components from the first 
dimension and $12$ components from the second dimension of the observations (at random observation times). 
The average time between each data point was taken as 1,  so as to reflect low-frequency observations.
In the PF,   the number of particles is set to be $\mathcal{O}(T)$,  which often stabilizes the variance of the likelihood estimator.  We then run the PF with the EM discretization method and the PF using the DB approach, each for $100$ times, varying the discretization levels from $2$ to $8$. We examine the performance of the estimated likelihood variance, which is displayed in Figure \ref{fig:likelihood_comp}.

The results show that as the discretization level is increased, the variance of the estimated likelihood from the PF using the EM method explodes, while the variance from the PF using the DB approach remains stable and relatively low.
This demonstrates the limitations of adopting the EM discretization method when dealing with low-frequency data, compared to the more robust DB approach.

\subsection{MLPMCMC Performance and Numerical Rates}

To evaluate the cost rates of our multilevel Particle MCMC (MLPMCMC) algorithm, we generate $T = 65$ observations for the same bi-variate OU process as in the previous Section.  For each dimension, we omit $16$ observations at random and the average time between each data point was taken as 1.
The parameterization for the diffusion coefficient $\Sigma$ is taken as:
$$
\Sigma = \begin{bmatrix}
    \sigma^2_1, & \rho \sigma_1 \sigma_2 \\
    \rho \sigma_1 \sigma_2 & \sigma^2_2
\end{bmatrix}.
$$
To set the priors all the parameters $\sigma_1,\sigma_2,\rho,A_{11},\dots,A_{22}$ are chosen as independent. The 
parameters in $\Sigma$ are such that $\log(\sigma_1), \log(\sigma_2)$ are Gaussian and $\log\left(\{\rho+1\}/\{\rho-1\}\right)$ is also Gaussian.
 $A_{11},\dots,A_{22}$ are Gaussian as well.  In the PFs that are to be used,  the number of particles is set to be $\mathcal{O}(T)$.
For the DB approach the auxiliary process is $d\tilde{X}_t=\Sigma dW_t$,  which allows one to compute
all required quantities.  The transition density $\hat{f}_{\theta}(x_{t_i}|x_{t_{i-1})}$ is bi-variate
Gaussian with mean $x_{t_{i-1}}$ and covariance matrix that is $0.01\Sigma^2(t_i-t_{i-1})$.
To evaluate the performance of the MLPMCMC sampler, we set the minimum level 
of time discretization of be 5,  written $L_{\min}=5$.

%In one setting, we fix the diffusion coefficient $\Sigma$ and estimate the four components $A_{11}, A_{12},A_{21},A_{22}$ of the drift matrix $A$. In a separate setting, we fix the drift matrix $A$ and estimate the three parameters $\sigma_1, \sigma_2$ and $\rho$  

\begin{figure}[htbp]
    \centering
    \includegraphics[scale=0.35, trim={2.5cm 3cm 2.5cm 3cm},clip]{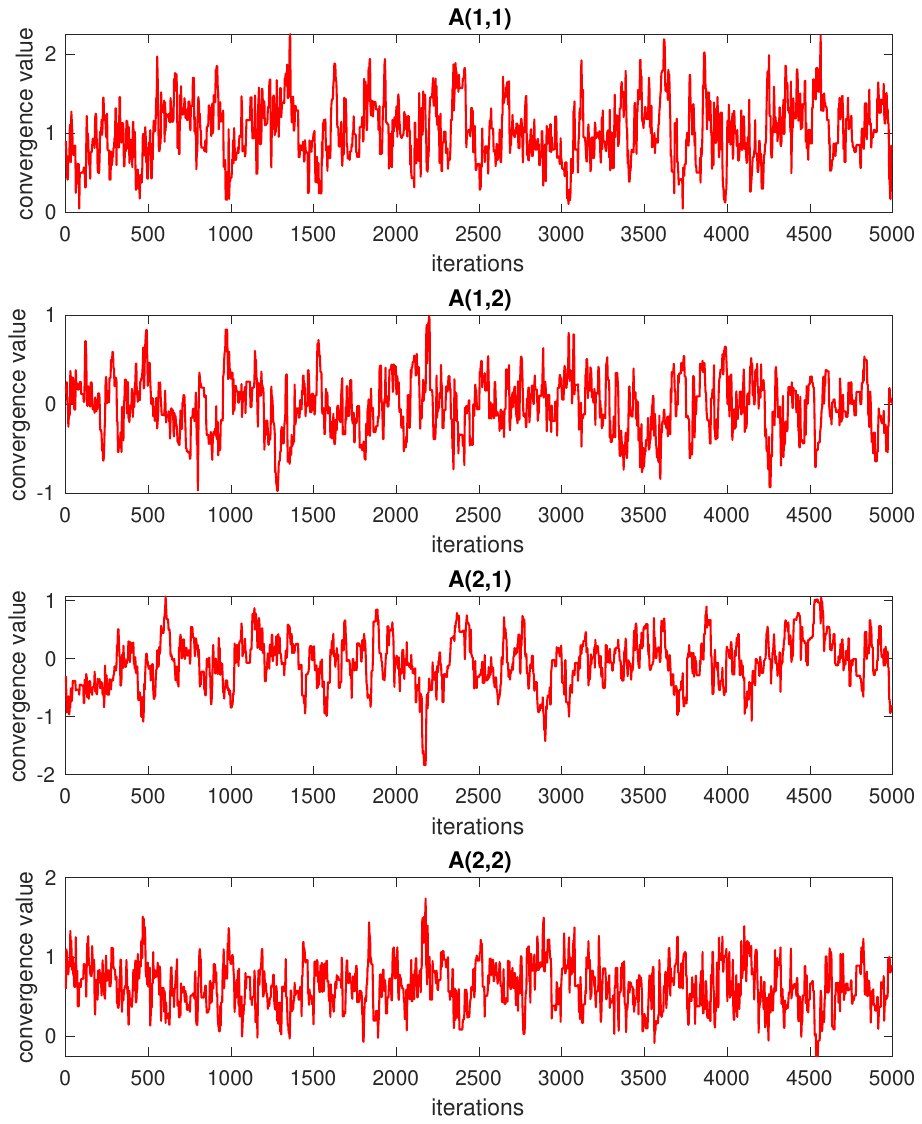}
    \includegraphics[scale=0.35, trim={2.5cm 3cm 2.5cm 3cm},clip]{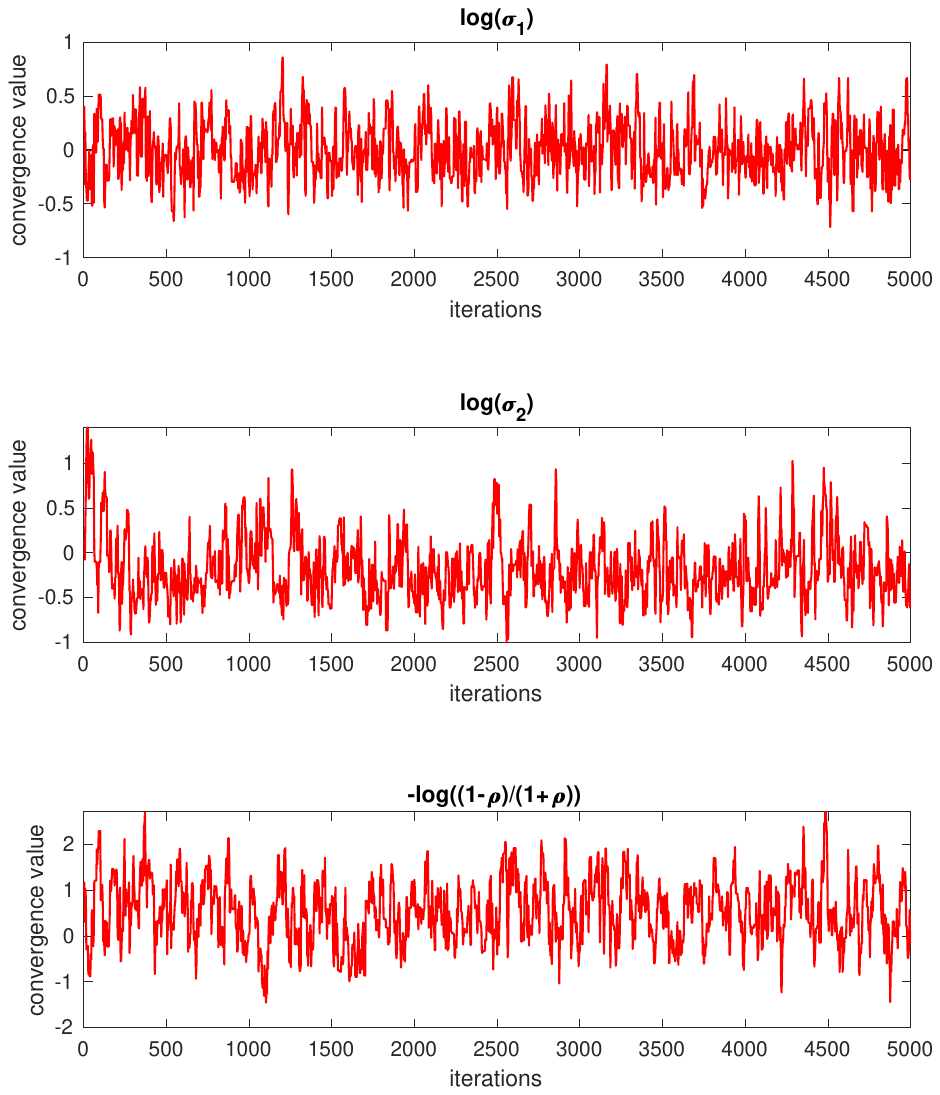}
    \caption{The convergence plot of the MLPMCMC for the drift coefficient A (left) and diffusion coefficient $\Sigma$ (right). This is for the OU model.}
    \label{fig:MLPMCMC_conv}
\end{figure}

In Figure \ref{fig:MLPMCMC_conv} we display the convergence of the MLPMCMC algorithm when considering
the samples at level 7 and over 10000 iterations.  In this example,  the MCMC performs adequately in terms of its mixing performance.
To further investigate the relationship between computational cost and mean squared error (MSE) as discussed previously, we test the estimated rate of decrease of the MSE as the cost increases on a logarithmic scale. The results are presented in Table \ref{tab:rate_BiOU}. For our multilevel method, the expected rate of decrease should be around $-1.0$ to $-1.1$ for each parameter. The results in the table are consistent with this expectation, verifying the conjectures that were made previously in the article.

%These findings demonstrate the efficiency and effectiveness of the MLPMCMC algorithm in estimating the model parameters for the non-synchronously bi-variate OU process, with the cost-MSE relationship aligning with the theoretical properties of the multilevel approach.

\begin{table}[h!]
\caption{Numerical Rates of the drift and diffusion coefficients for simulated OU process}
\centering
\begin{tabular}{cc}
  \begin{tabular}{c|c}
    Drift & Rates \\
    \hline
    $A_{11}$ & $-1.23$ \\
    $A_{12}$ & $-1.26$ \\
    $A_{21}$ & $-1.22$ \\
    $A_{22}$ & $-1.17$ \\
  \end{tabular}
  &
  \begin{tabular}{c|c}
    Diffusion & Rates \\
    \hline
    $\sigma_1$ & $-1.21$ \\
    $\sigma_2$ & $-1.23$ \\
    $\rho$ & $-1.36$ \\
  \end{tabular}
\end{tabular}
\label{tab:rate_BiOU}
\end{table}

\subsection{Stochastic Lotka--Volterra Predator-Prey Model}

We apply our method to the 
stochastic Lotka--Volterra predator-prey model given by:
\begin{eqnarray*}
dX_t^1 & = & X_t^1 (\alpha  - \beta X_t^2)dt + \sigma_1 X_t^1 dW^{1}_t \\
 dX_t^2 & = & X_t^2 (\zeta X_t^1 - \gamma)dt + \sigma_2 X_t^1 dW^{2}_t
\end{eqnarray*}
In this model, $X_t^1$ is the prey population (e.g.~rabbits) and $X_t^2$ is the predator population (e.g.~foxes) at time $t$. 
The parameter $\alpha$ is the intrinsic growth rate of the prey, while $\beta$ represents the predation rate coefficient. The contribution efficiency $\zeta$ indicates how prey presence affects predator growth, and $\gamma$ is the natural death rate of predators. Parameters $\sigma_1$ and $\sigma_2$ reflect the intensity of environmental fluctuations for prey and predators, respectively.   All of the unknown parameters have independent Gaussian priors,  after being logarithmically transformed onto the real line.
%$W^1_t$ and $W^2_t$ are correlated Brownian motions with correlation parameter $\rho$, representing random disturbances to both populations.  

We will consider using two real datasets. The first dataset comprises wildebeest and zebra populations from \cite{fay2006lion}.  To optimize model fitting, we rescaled the data by dividing both columns by 1000 and adjusted the start time to 0. We then omitted portions of the observations for each species, resulting in a final dataset which is available on request.  The second dataset consists of historical data for two competitors in the Greek cell phone telecommunications market, extracted from \cite{fay2006lion} (specifically, the first two columns of Table 1). To enhance model fitting, we rescaled the data to commence from time 0 and omitted portions of the observations, leading to a final dataset which is again available on request.

The transition density $\hat{f}_{\theta}(x_{t_i}|x_{t_{i-1})}$ is taken as (conditionally) independent in each dimension and log normals with location in dimension $j\in\{1,2\}$ as $\log(x_{t_{i-1}}^j)-\sigma_j^2/2$ and scale $\sigma_j$.  For the diffusion bridge that is used inside the particle filter,  we need to specify the auxiliary process,
which we shall give on an interval $[0,t_1]$.  Consider starting at $(x^1,x^2)$ with the end point of the bridge
$((x')^1,(x')^2)$, then we use the following process:
\begin{align*}
    d \tilde{X}_{t}^{1} &= \tilde{X}_{t}^{1} \left[ (\alpha - \beta x^{2})\left(1 - \frac{t}{t_1}\right) + (\alpha - \beta (x')^{2})\frac{t}{t_1} \right] dt + \sigma_1 \tilde{X}_{t}^{1} dW^1_t \\
    d \tilde{X}_{t}^{2} &= \tilde{X}_{t}^{2} \left[ (\zeta x^{1} - \gamma)\left(1 - \frac{t}{t_1}\right) + (\zeta (x')^{1} - \gamma)\frac{t}{t_1} \right] dt + \sigma_2 \tilde{X}_{t}^{2} dW^2_t.
\end{align*}
The transition density is available as the SDE can be explicitly solved.

\begin{figure}[htbp]
    \centering
    \vspace{-10pt}
    \hspace{-20pt}
    \includegraphics[scale=0.35, trim={2.5cm 3cm 2.5cm 3cm},clip]{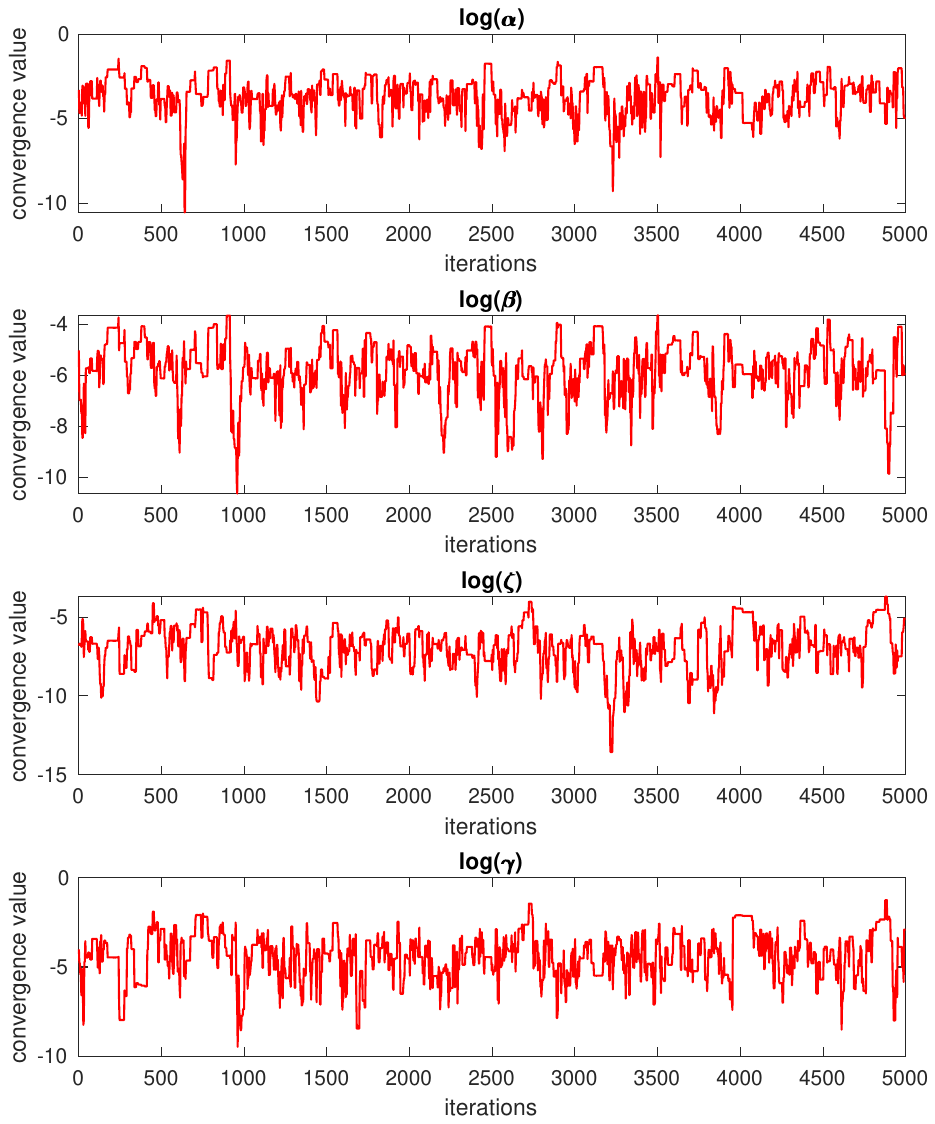}
    \includegraphics[scale=0.35, trim={2.5cm 3cm 2.5cm 3cm},clip]{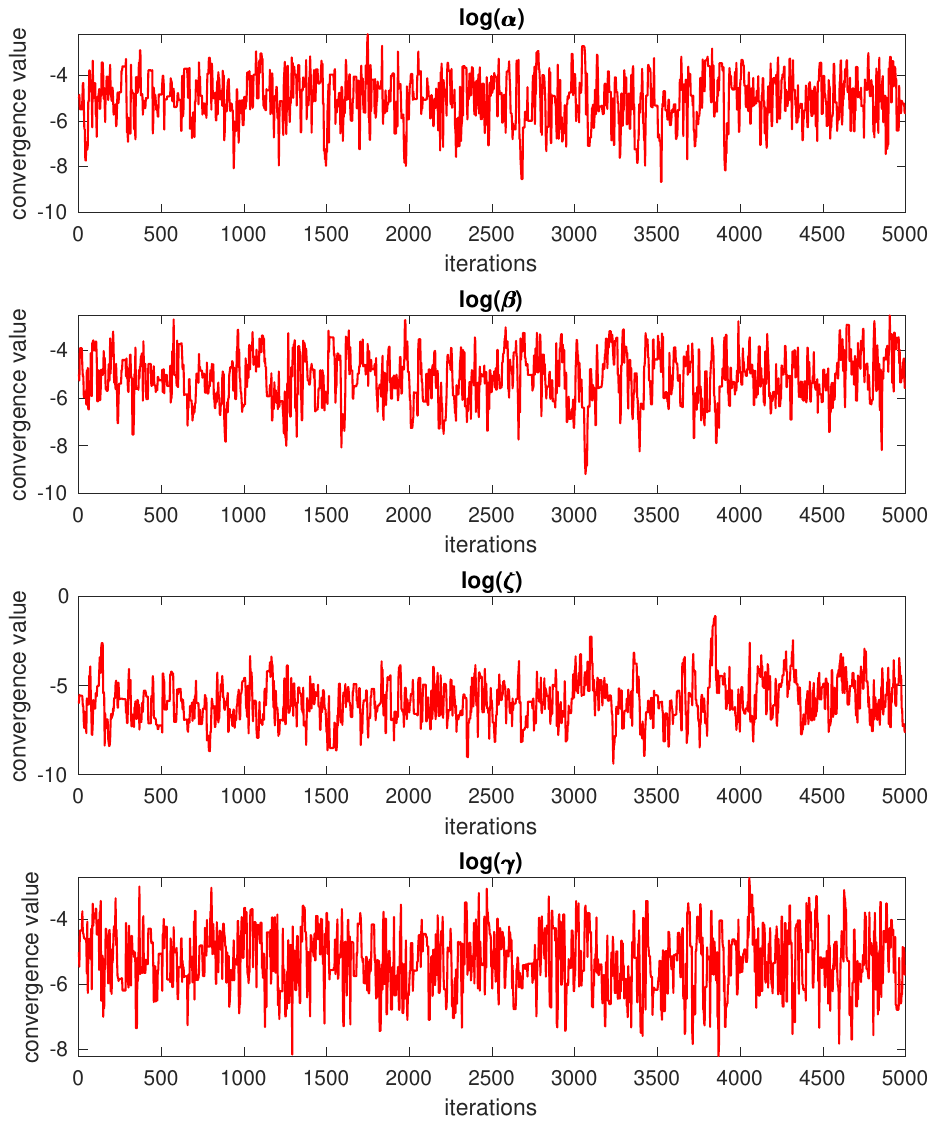}
    \caption{Convergence Plots for the Drift Parameters for the Two Data Sets for the Stochastic Lotka--Volterra Model.  The right column is the cell phone data set.}
    \label{fig:bz_ml}
\end{figure}

To evaluate the performance of the MLPMCMC sampler on the two real datasets mentioned earlier, we set a minimum level of ($L_{\min} = 4$) and consider convergence at discretization at level 6 which is the maximum level considered.   We present the convergence in terms of the simulated drift parameters in Figure \ref{fig:bz_ml} for both data sets.  The algorithm appears to perform reasonably well over a short run,  which is all that is needed at the highest level.  Similar performance was found at the lower levels.

To further investigate the relationship between computational cost and MSE for the PMCMC algorithm and the MLPMCMC method, we evaluate the estimated MSE reduction rate as drift parameter costs increase on a logarithmic scale. Table \ref{tab:rates_comp_slv} summarizes the findings, indicating an expected theoretical reduction rate of approximately $-1.2$ to $-1.3$ per parameter for the multilevel method and around $-1.5$ for the single level. The empirical results corroborate these expectations, validating the earlier conjectures presented in the article.

\begin{table}[h!]
\caption{Drift Coefficient Rates for PMCMC and MLPMCMC. This is for the Stochastic Lotka--Volterra Model and the cell phone data set.}
\centering
\begin{tabular}{cc}
  \begin{tabular}{c|c}
    PMCMC & Rates \\
    \hline
    $\log(\alpha)$ & $-1.53$ \\
    $\log(\beta)$ & $-1.54$ \\
    $\log(\gamma)$ & $-1.50$ \\
    $\log(\zeta)$ & $-1.51$ \\
  \end{tabular}
  &
  \begin{tabular}{c|c}
    MLPMCMC & Rates \\
    \hline
     $\log(\alpha)$ & $-1.36$ \\
    $\log(\beta)$ & $-1.32$ \\
    $\log(\gamma)$ & $-1.42$ \\
    $\log(\zeta)$ & $-1.30$ \\
  \end{tabular}
\end{tabular}
\label{tab:rates_comp_slv}
\end{table}

\end{document}